\documentclass[preprint]{aastex63}

\usepackage{natbib}
\usepackage{epsfig}
\usepackage{graphicx}
\usepackage{multirow}
\usepackage[caption=false]{subfig}
\usepackage{epstopdf}
\usepackage{url}

\shorttitle{Parallaxes of M-dwarf Eclipsing Binaries}

\shortauthors{van Belle, Schaefer, \& von Braun et al.}

\begin{document}

\title{{\it HST}/FGS Trigonometric Parallaxes of M-dwarf Eclipsing Binaries}


\author{Gerard T. van Belle}
\affiliation{Lowell Observatory}
\author{Gail H. Schaefer}
\affiliation{The CHARA Array of Georgia State University}
\author{Kaspar von Braun}
\affiliation{Lowell Observatory}
\author{Edmund P.~Nelan}
\affiliation{Space Telescope Science Institute}
\author{Zachary Hartman}
\affiliation{Georgia State University}
\affiliation{Lowell Observatory}
\author{Tabetha S. Boyajian}
\affiliation{Louisiana State University}
\author{Mercedes Lopez-Morales}
\affiliation{Harvard-Smithsonian Center for Astrophysics}
\author{David R. Ciardi}
\affiliation{NASA Exoplanet Science Institute, California Institute of Technology}


\begin{abstract}

{\it Hubble Space Telescope} ({\it HST}) Fine Guidance Sensor (FGS) trigonometric parallax observations were obtained to directly determine distances to five nearby M-dwarf / M-dwarf eclipsing binary systems.
These systems are intrinsically interesting as benchmark systems for establishing basic physical parameters for low-mass stars, such as luminosity $L$, and radius $R$.
{\it HST}/FGS distances are also one of the few direct checks on {\it Gaia} trigonometric parallaxes, given the comparable sensitivity in both magnitude limit and determination of parallactic angles.  
A spectral energy distribution (SED) fit of each system's blended flux output was carried out, allowing for estimation of the bolometric flux from the primary and secondary components of each system.
From the stellar $M$, $L$, and $R$ values, the low-mass star relationships between $L$ and $M$, and $R$ and $M$, are compared against idealized expectations for such stars.
An examination on the inclusion of these close M-dwarf/M-dwarf pairs in higher-order common proper motion (CPM) pairs is analysed; each of the 5 systems has indications of being part of a CPM system.
Unexpected distances on interesting objects found within the grid of parallactic reference stars are also presented, including a nearby M dwarf and a white dwarf.

\end{abstract}

%




%
%
%


\section{Introduction}\label{sec_introduction}


We obtained {\it Hubble Space Telescope} ({\it HST}) Fine Guidance Sensor (FGS) trigonometric parallax observations to directly determine distances to five nearby M-dwarf / M-dwarf eclipsing binary systems. Empirical distances yield or place constraints on the luminosities of these systems and individual components and consequently on low-mass stellar models.  Fundamental astrophysical parameters of low-mass ($M < 1 M_{\odot}$) stars are of particular importance given that the large majority of the stars in the Galaxy are low-mass objects \citep{Turnbull2003ApJS..149..423T,Lopez-Morales2005ApJ...631.1120L,Henry2006AJ....132.2360H,Henry2007IAUS..240..299H}.  While the accuracy of theoretically determined astrophysical parameters of low-mass stars is continuously being improved, there is still considerable disagreement between models and observation for both single stars and components of binary systems \citep{Torres2002ApJ...567.1140T,Ribas2003A&A...398..239R,Berger2006ApJ...644..475B,Torres2010A&ARv..18...67T,Boyajian2012ApJ...757..112B,vonBraun2014MNRAS.438.2413V,Mann2015ApJ...804...64M,vonBraun2017arXiv170707405V}. Studying these objects is therefore crucial to understanding their intrinsic stellar astrophysics, the extent of their habitable zones, and the physics of the increasing number exoplanets being discovered about these low mass stars \citep[e.g.,][and references therein]{Dressing2015ApJ...807...45D,Ballard2016ApJ...816...66B,Ballard2019AJ....157..113B}.

Observationally, stellar luminosity is obtained from the combination of  distance and bolometric flux. Eclipsing binaries are ideal candidates to reconcile those parameters predicted by models with the observational quantities. However, only a small number ($N \lesssim 10$) of nearby ($d<$150pc) eclipsing M-dwarf binaries are known \citep{Lopez-Morales2007ApJ...660..732L}, and their distance determinations are either poorly constrained or entirely unknown. Thus, even with known bolometric fluxes for these systems, their luminosities remain model dependent.

Between August 2005 and May 2009, {\it HST} was operating using only two of its originally six gyros\footnote{See \url{https://www.nasa.gov/home/hqnews/2005/aug/ HQ\_05242\_hst\_2\_gyros.html} and \url{https://www.nasa.gov/mission\_pages/ hubble/servicing/index.html}.}. In 2007--2009, we observed the well-studied systems GU Boo, CM Dra, YY Gem, and the more recently discovered binaries NSVS01031772 \citep[``NSVS0103'' hereafter]{Lopez-Morales2006astro.ph.10225L} and TrES-Her0-07621 \citep[``TrES-Her0'' hereafter]{Creevey2005ApJ...625L.127C} using FGS (see Table \ref{tab_FGS_obslog}). Both respective eclipsing components of each of these binary systems are M dwarfs. Two other nearby systems, NSVS07394765 \citep{Coughlin2007JSARA...1....7C} and CU Cnc \citep{Delfosse1999A&A...341L..63D}, unfortunately did not present enough observing opportunities in {\it HST's} two-gyro mode for parallax determination with FGS.

Of these seven systems, only CU Cnc is found in the Hipparcos catalog, with a 5.67 mas parallax error \citep{Perryman1997A&A...323L..49P}; YY Gem can be found in the Yale GCTP with a 2.5 mas parallax error \citep{vanAltena1995yCat.1174....0V}; the other five systems had no trigonometric parallaxes available, clearly illustrating the paucity of distance information for these systems at the time.  This has more recently been addressed via a recent FGS study by \citet{Benedict2016AJ....152..141B} measuring trigonometric parallaxes of M-dwarf binaries, and, of course, the recent release of {\it Gaia} Data Release 2 \citep[DR2;][]{Gaia2018A&A...616A...1G} provides an interesting astrometric comparison data set relative to our {\it HST}/FGS results.  These observations -- selected for {\it HST} orbits long before {\it Gaia} flew -- provide a compelling cross-check on {\it Gaia} results.  While that mission has proven itself to be a superlative source of distances and other astrophysical data, it is not infallible - for example, the {\it Gaia} DR2 distance on benchmark star RR Lyr is spurious with a value of $\pi = -2.61 \pm 1.15$.


In order to calculate stellar luminosities, we supplement our direct distances based on trigonometric parallaxes with empirical estimates of the bolometric fluxes of the binary systems, as well as the bolometric fluxes of the respective individual stellar components, based on spectral energy distribution (SED) fitting of spectral templates to broad-band and narrow-band photometry. An ancillary product of this approach is an estimate of the angular stellar diameter for every stellar component of every system, which provide a sanity check to their counterparts produced by the studies of the eclipsing binary systems. Finally, all of our targets have dynamically determined masses with errors less than $0.5$\%, which, coupled with directly determined distances as model-independent constraints, provide important anchors for the low mass end of the mass-luminosity relationship.


We describe our observations in \S \ref{sec_observations}, which include the {\it HST}/FGS observations and the selection of astrometric reference stars with respect to which trigonometric parallaxes of targets are measured (\S \ref{sec_HSTobservations}), plus spectroscopy and photometry for our targets and astrometric reference stars (\S \ref{sec_spectroscopy} and \ref{sec_photometry}). \S \ref{sec_analysis} describes our data analysis, including SED fitting (\S \ref{sec_sedFit}), the determination of the bolometric flux values of the individual components of our target systems (\S \ref{sec_binarysed}),  distance calculations to the reference stars (\S \ref{sec_inferred_distances}) and target stars (\S \ref{sec_trig_parallax}), and comparisons with $Gaia$ values (\ref{sec_compare_gaia}).
Serendipitous findings within our data can be found in \S \ref{sec_serendipity}, such as the direct distance determination to a white dwarf and another M-dwarf.  The distance-enabled astrophysics is presented in \S \ref{sec_discussion}.
We conclude in \S \ref{sec_conclusion}.


\section{Observations} \label{sec_observations}

Parallax observations can be subdivided into observations from global astrometry (i.e. {\it Gaia}), and small- or single-field astrometry (i.e. {\it HST}/FGS);
for an expanded discussion comparing two approaches, see the discussion in \S 2 of \citet{Benedict2017PASP..129a2001B} and references therein.
The {\it HST}/FGS method of trigonometric parallax is based on measuring relative positions of foreground target stars with respect to background reference stars at different phase angles during the Earth's motion around the Sun.
In order to convert these data to absolute distances, we estimate the distances to the background stars by collecting ancillary photometric and spectroscopic data.  With spectral types and luminosity classes determined from the spectra, we use photometry to construct spectral energy distributions to determine interstellar reddening, absolute magnitude, and hence distance, for each reference star.  This allows us to correct the positions of the reference stars measured by the FGS for residual astrometric motion caused by their parallaxes; thereby converting the distances measured for the M-dwarf eclipsing binaries to an absolute value.  These distance estimates and final astrometric model are computed and presented in \S \ref{sec_analysis}.


\subsection{{\it HST}/FGS Observations}\label{sec_HSTobservations}

The {\it HST}/FGS instrument is described in \citet{Nelan2012fgsi.book.....N}, and an overview of results from the FGS is presented by \citet{Benedict2017PASP..129a2001B}.  The astrometric data from this program are available through the \href{http://www.stsci.edu/hst/scheduling/program_information}{HST Program Information website} for program 11213.

\begin{figure*}
\plotone{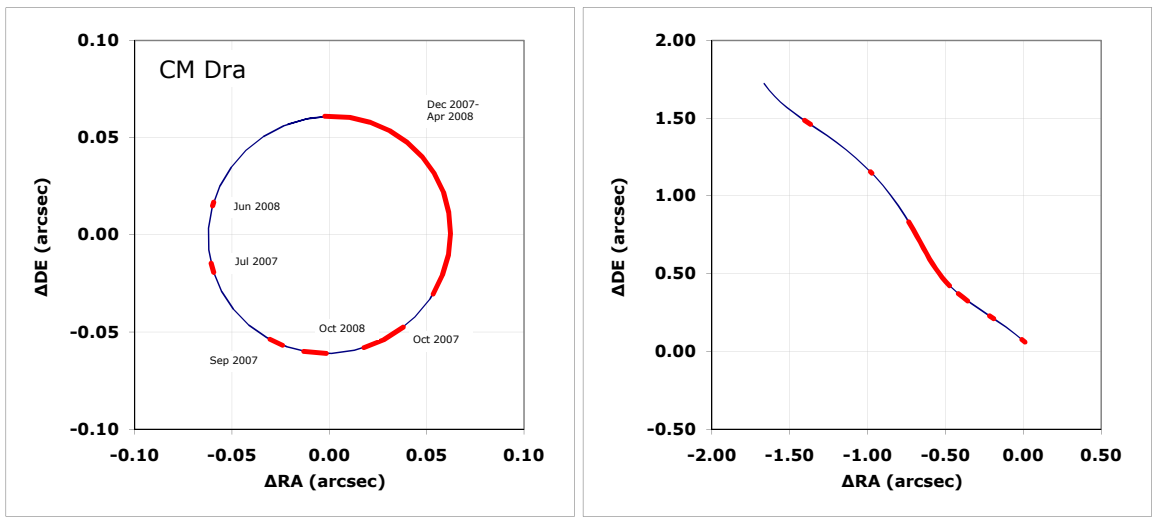}
\caption{\label{fig_CMDra} Right panel: Observed motion of the system on the sky, which includes the effects of parallax and proper motion.  Left panel: Parallactic ellipse expected for the CM Dra system, isolated from proper motion effects.  In both panels, the {\it HST}/FGS epochs of observation are indicated by the red points. For more details, see \S\ref{sec_HSTobservations}.
\\ \\}
\end{figure*}

The target and reference stars were observed following the methodology described in \citet{Benedict2007AJ....133.1810B}, incorporating seven epochs of observations on each object spread over 1.5 to 2.0 years (Table \ref{tab_FGS_obslog}).
During our observations, FGS-1r was in POS mode using the astrometrically calibrated F583W filter.  Multiple observations were obtained of the target and the reference stars available in the field during each {\it HST} orbit.  All targets except TrES-Her0 are brighter than $V=14$, requiring minimal (1.2 minute) exposure time and overhead per target. The exposure and overhead requirements for TrES-Her0 were still small ($<$5 minutes), compared to a nominal 53 minute window of observations during each orbit. Data sets were taken at epochs closest to maximum parallactic excursion, with sets at the available far ends of the Cycle calendar window (July 2007 - December 2009) in order to disentangle parallax and proper motion effects and exclude times of eclipse of the binary system.  An example of expected parallactic excursion on the sky is plotted for CM Dra in Figure \ref{fig_CMDra}.

\begin{figure}
  \includegraphics[clip,width=0.66\columnwidth,angle=90]{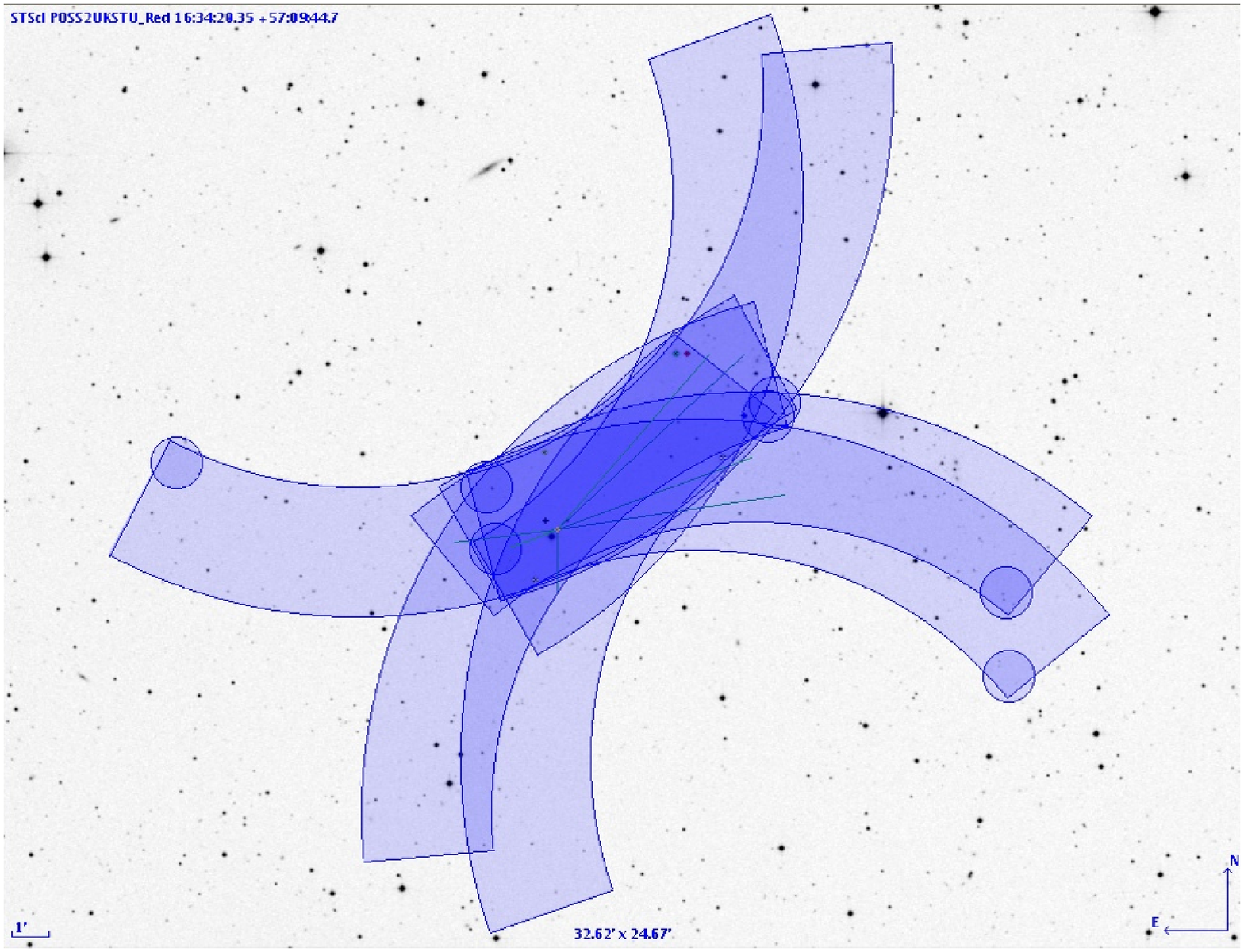} \\%
  \includegraphics[clip,width=0.75\columnwidth]{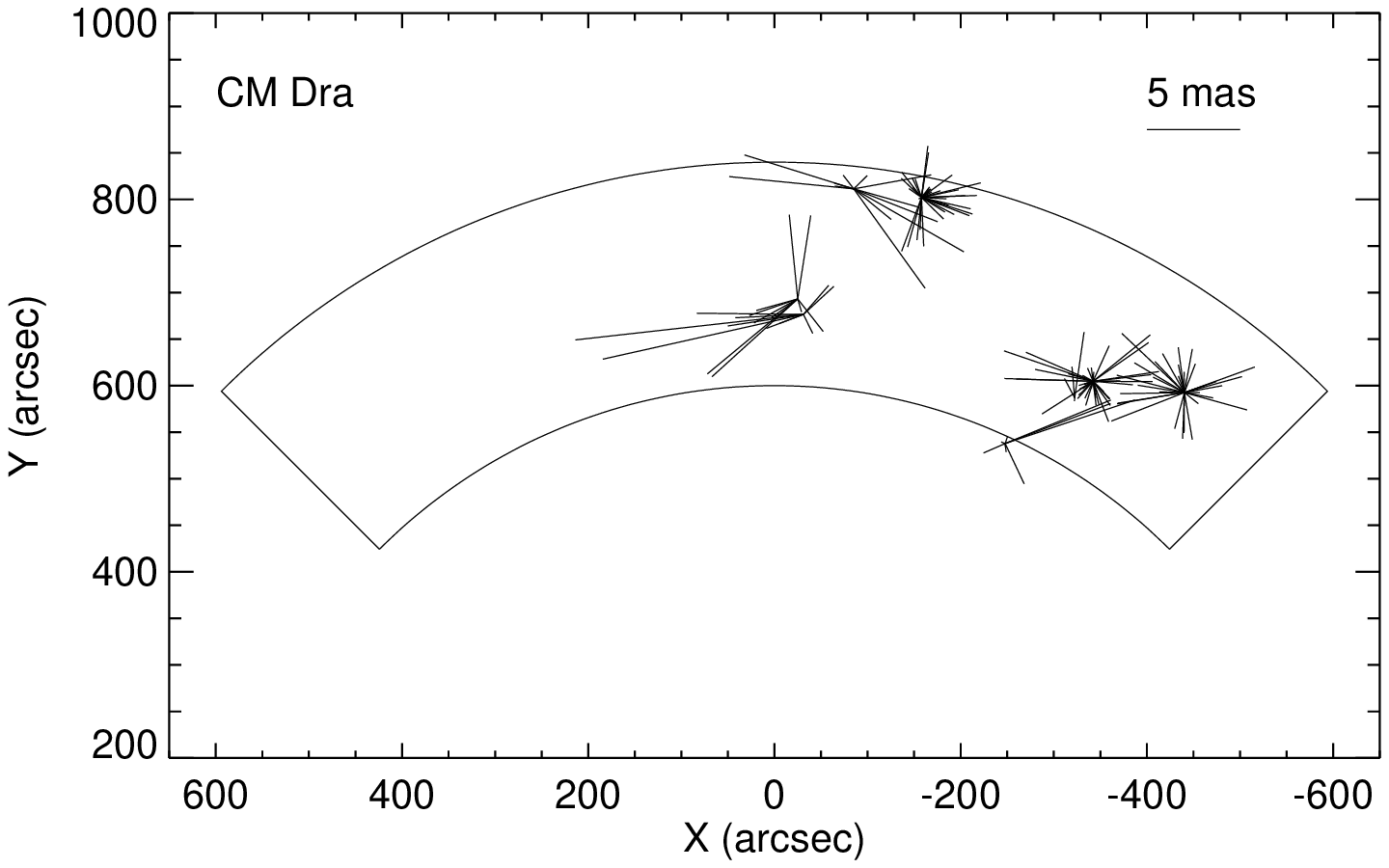}%
\caption{`Pickle plots' of the {\it HST}/FGS field-of-view (FOV). Top: Multiple epochs of on-sky projections of the {\it HST}/FGS FOV for CM Dra. Bottom: The reference epoch for the astrometric fit for the CM Dra field showing the residuals in the positions of target and reference stars over the course of the FGS observations. For more details, see \S\ref{sec_HSTobservations}. \\ \label{fig_pickle_plots}}
\end{figure}

During the course of the observations, {\it HST} was restricted to two-gyro pointing mode.  The impact of this restriction was that roll angles and times of observations for targets at particular sky locations were limited to certain values or not possible at all.  As noted in Section \ref{sec_introduction}, two otherwise appealing targets for this study (CU Cnc and NSVS07394765) were unavailable due to this mode.  For the available targets, restrictions in roll angle limited our options for selection of parallactic background reference stars, which required judicious scheduling of FGS observations, as we illustrate in Figure \ref{fig_pickle_plots}.  Because of the limitations on the roll angle, sometimes only a subset of the reference stars could be observed during a given epoch.



The FGS data were reduced and calibrated as described in \citet{McArthur2001ApJ...560..907M}, \citet{Benedict2002AJ....123..473B,Benedict2002AJ....124.1695B,Benedict2007AJ....133.1810B}, and \citet{Soderblom2005AJ....129.1616S}; a more recent investigation using FGS is presented in some detail in \citet{Chaboyer2017ApJ...835..152C}.  In general, the FGS reduction pipeline extracts the position of each star observed in the field and applies corrections for geometric distortion, differential velocity aberrations, spacecraft jitter, and drift.  These positions are used as input to the astrometric model discussed in Section~\ref{sec_trig_parallax}.




\clearpage
\begin{deluxetable}{lll}




\tablecaption{FGS Observing Log.\label{tab_FGS_obslog}}


\tablehead{\colhead{Target} & \colhead{UT Date} & \colhead{UT Time}}  
\startdata
CM DRA    &  2007-09-07 &  01:12 \\
          &  2008-01-02 &  05:30 \\
          &  2008-02-24 &  04:24 \\
          &  2008-06-15 &  19:15 \\
          &  2008-07-07 &  07:13 \\
          &  2008-12-17 &  03:13 \\
          &  2009-09-20 &  03:24 \\
GU Boo    &  2007-12-08 &  01:03 \\
          &  2008-01-12 &  00:13 \\
          &  2008-03-26 &  20:44 \\
          &  2008-05-22 &  16:11 \\
          &  2008-07-04 &  13:36 \\
          &  2008-12-06 &  03:24 \\
          &  2008-12-21 &  18:57 \\
NSVS0103  &  2007-10-29 &  01:23 \\
          &  2008-01-15 &  01:58 \\
          &  2008-05-03 &  15:19 \\
          &  2008-06-12 &  08:40 \\
          &  2008-08-06 &  00:51 \\
          &  2008-09-21 &  01:37 \\
          &  2008-12-20 &  02:59 \\
TRES-HER0 &  2007-10-21 &  03:37 \\
          &  2008-01-02 &  00:04 \\
          &  2008-03-23 &  14:30 \\
          &  2008-07-01 &  12:11 \\
          &  2008-09-25 &  12:58 \\
          &  2009-01-02 &  20:11 \\
YY GEM    &  2007-11-01 &  22:00 \\
          &  2007-12-15\tablenotemark{a} &  19:26 \\
          &  2008-02-05 &  16:41 \\
          &  2008-09-15 &  12:49 \\
          &  2008-09-15 &  20:49 \\
          &  2008-11-02 &  13:15 \\
          &  2008-12-13 &  16:54 \\
\enddata
\tablecomments{For more details, see \S\ref{sec_HSTobservations}.}
\tablenotetext{a}{Target acquisition failed for the observations of the YY Gem field on 2007-12-15.}




\end{deluxetable}


\subsection{WIYN HYDRA Spectra} \label{sec_spectroscopy}

To classify the spectral types and luminosity classes of the reference stars, we used the Hydra multi-fiber spectrograph on the WIYN 3.5~m telescope at Kitt Peak.  We used the blue fiber cable with the BG-39 filter and a grating of 600 lines mm$^{-1}$ blazed at 13.9$^{\circ}$ in the second order.  This setup provided spectra from 3800 to 5200~\AA\, at a dispersion of 0.70~\AA\,pixel$^{-1}$ with a resolving power of 1,581.  This spectral region has numerous lines that are useful for determining spectral types and luminosity classifications (e.g. Gray's Digital Classification Atlas\footnote{http://nedwww.ipac.caltech.edu/level5/Gray/frames.html}).  We recorded the data using the Bench Spectrograph Camera.

The Hydra spectra were obtained on UT 2010 April 23-26.  We acquired spectra on all of the reference stars using one to two fiber positioning setups per FGS field.  Total integration times per star ranged from 10 to 40 minutes.  In each field we assigned 13$-$19 sky fibers for removing the background sky.  We also obtained dome flats in the blue circle fiber configuration and bias frames.  For wavelength calibration, we acquired CuAr lamp spectra for each fiber configuration.  The spectra were flat-fielded, extracted, wavelength calibrated, and sky subtracted using the IRAF Hydra package.

\begin{figure*}
\epsscale{0.90}
\plotone{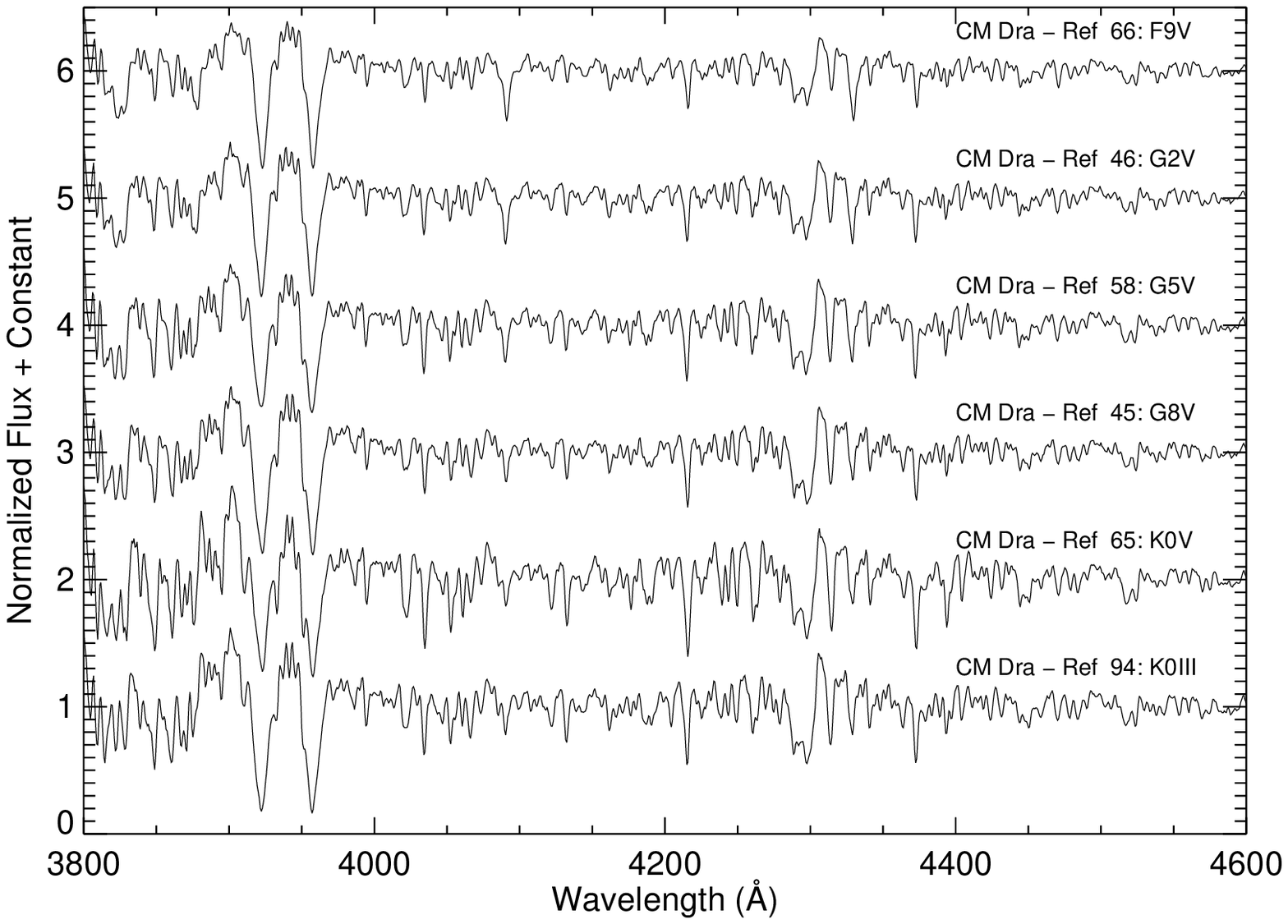}
\caption{\label{fig_refstars_spectra_cmdra} Reference star spectra for CM Dra. For more information, see \S \ref{sec_spectroscopy}.}
\end{figure*}

\begin{figure*}
\epsscale{0.90}
\plotone{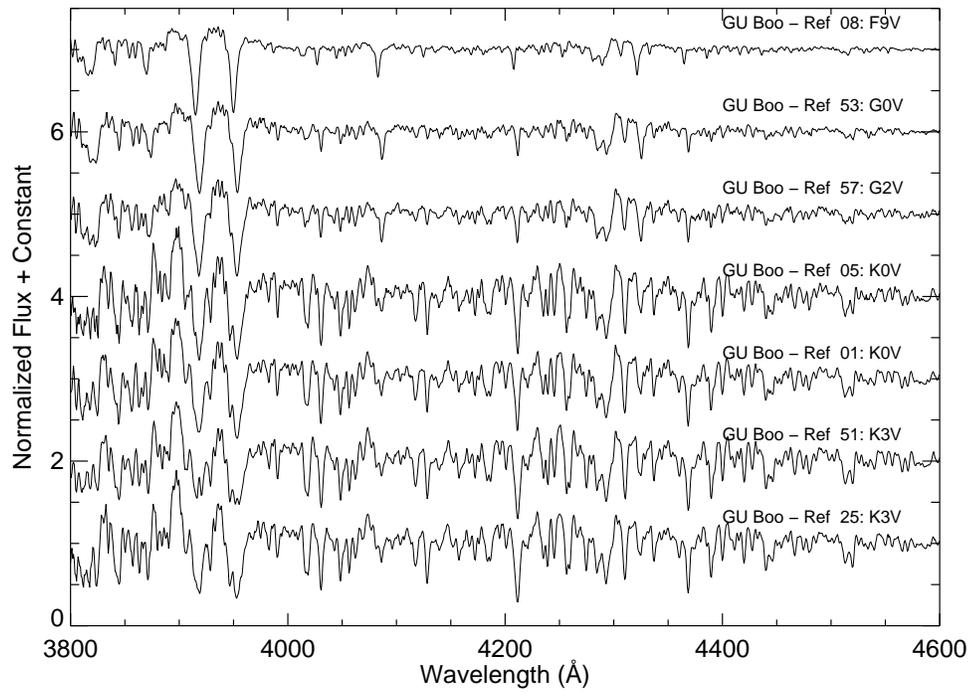}
\caption{\label{fig_refstars_spectra_guboo} Reference star spectra for GU Boo. For more information, see \S \ref{sec_spectroscopy}.}
\end{figure*}

\begin{figure*}
\epsscale{0.90}
\plotone{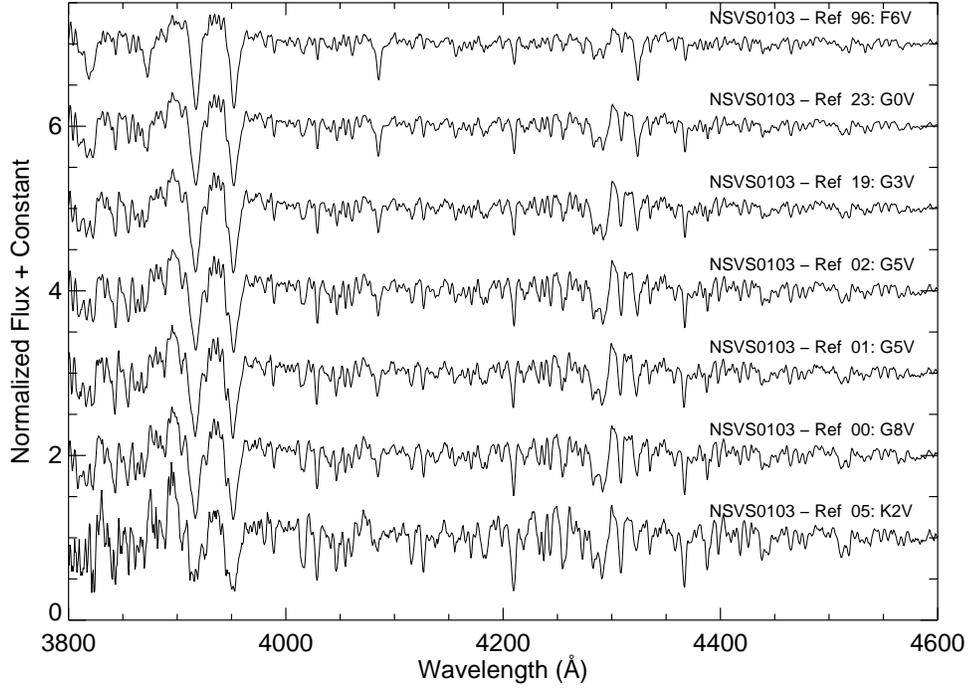}
\caption{\label{fig_refstars_spectra_nsvs0103} Reference star spectra for NSVS0103. For more information, see \S \ref{sec_spectroscopy}.}
\end{figure*}

\begin{figure*}
\epsscale{0.90}
\plotone{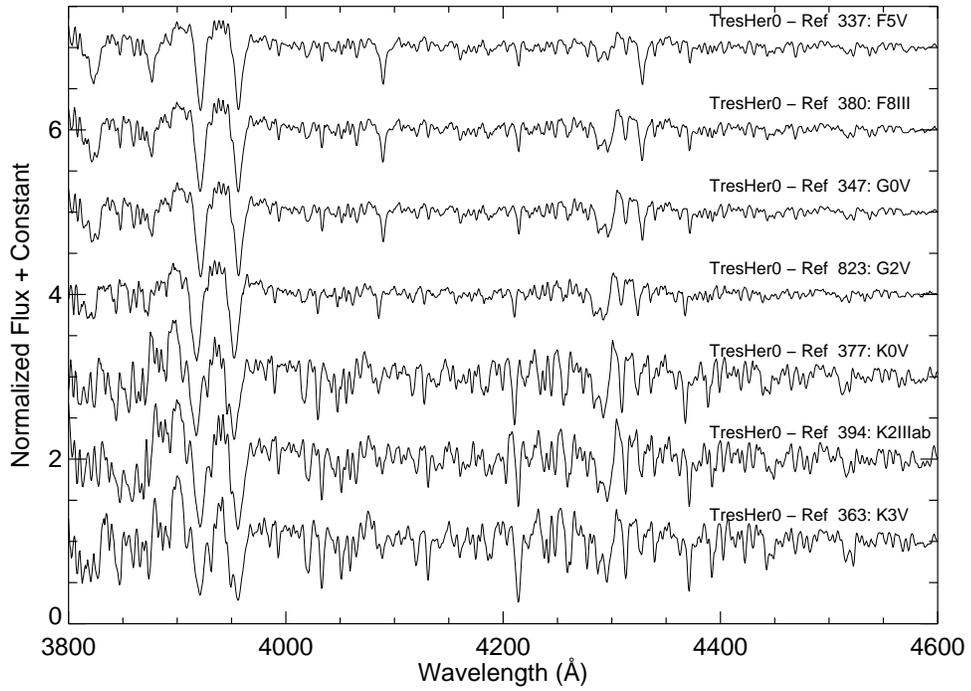}
\caption{\label{fig_refstars_spectra_tresher0} Reference star spectra for TrES-Her0. For more information, see \S \ref{sec_spectroscopy}.}
\end{figure*}

\begin{figure*}
\epsscale{0.90}
\plotone{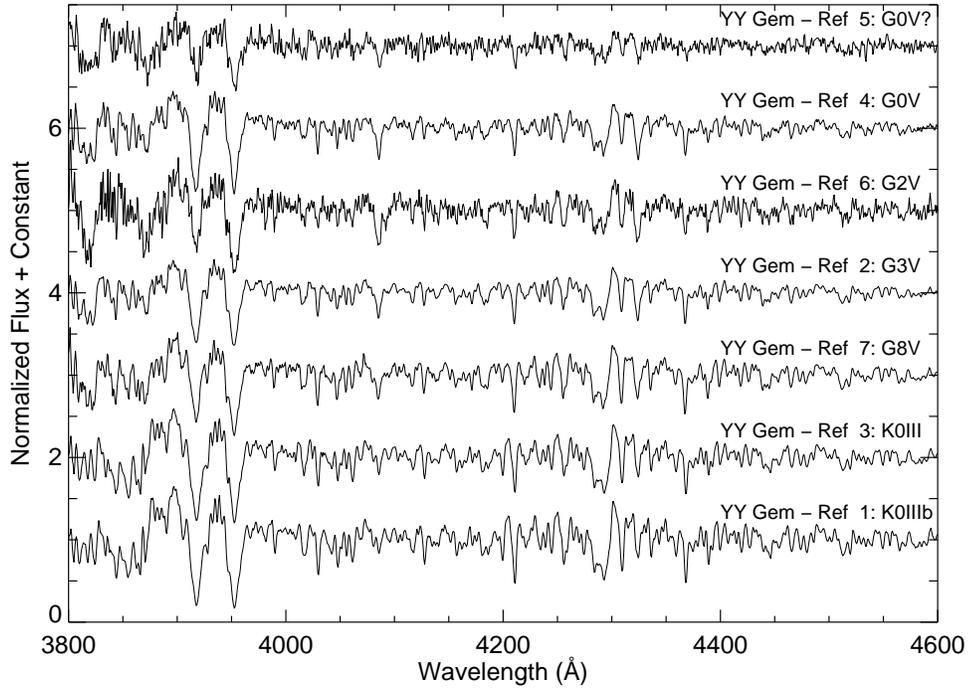}
\caption{\label{fig_refstars_spectra_yygem} Reference star spectra for YY Gem. For more information, see \S \ref{sec_spectroscopy}.}
\end{figure*}

We flattened the spectra using a spectral rectification technique that removes the shape of the stellar continuum, interstellar reddening, atmospheric extinction, and instrument response \citep{LaSala1985PASP...97..605L}.  In this procedure, we took the Fourier transform of every spectrum, padded the ends by applying a cosine bell apodization, and put it through a low-pass filter to remove the high frequency Fourier components.  We then transformed this spectrum back to get the underlying shape of the continuum and divided it out from the original spectrum to flatten it.  Spectral classification was performed by comparing relative line strengths of the reference stars with spectral standards observed with the same instrument by H. E. Bond \citep[e.g.,][]{Bond2013ApJ...765L..12B}.  The spectral templates included A, F, G, K, and M-type stars of main sequence, sub-giant, and giant classes with known spectral types.  We also observed a few more spectral standard stars during our run with Hydra to cover a broader range of spectral sub-types.  The spectral types and luminosity classes determined for the FGS reference stars are listed in Table \ref{tab_photometry}.  Plots of the spectra are shown in Figures \ref{fig_refstars_spectra_cmdra}--\ref{fig_refstars_spectra_yygem}.


\begin{deluxetable}{lccccccccccl}
\rotate 




\tablecaption{\label{tab_photometry}Target and reference star coordinates, photometry, and inferred spectral type: Johnson UBVRI and 2MASS JHKs.}


\tabletypesize{\tiny}
\tablehead{\colhead{Program ID} & \colhead{RA} & \colhead{DE}
& \colhead{U}& \colhead{B}& \colhead{V}& \colhead{R}
& \colhead{I} & \colhead{J} & \colhead{H} & \colhead{Ks} & \colhead{Spectral}\\
\colhead{} & \colhead{(hms)} & \colhead{(dms)} & \colhead{(mag)} & \colhead{(mag)} & \colhead{(mag)} & \colhead{(mag)} & \colhead{(mag)} & \colhead{(mag)} & \colhead{(mag)} & \colhead{(mag)} & \colhead{Type} }

\startdata
CM-DRA  &  16 34 20.417  &  +57 09 43.93  & \ldots & $14.45 \pm 0.01$ & $12.92 \pm 0.00$ & $11.63 \pm 0.01$ & $9.99 \pm 0.04$ & $8.50 \pm 0.02$ & $8.04 \pm 0.03$ & $7.80 \pm 0.02$ &  M4.5V \\
CM-DRA-REF45  &  16 33 57.786  &  +57 14 23.13  & $15.55 \pm 0.07$ & $14.23 \pm 0.01$ & $13.41 \pm 0.01$ & \ldots & \ldots & $11.76 \pm 0.02$ & $11.32 \pm 0.03$ & $11.21 \pm 0.02$ &  G8V?  \\
CM-DRA-REF46  &  16 33 55.605  &  +57 14 23.16  & $14.93 \pm 0.07$ & $14.93 \pm 0.00$ & $14.24 \pm 0.01$ & \ldots & \ldots & $13.08 \pm 0.02$ & $12.79 \pm 0.03$ & $12.75 \pm 0.03$ &  G2V  \\
CM-DRA-REF47  &  16 34 22.591  &  +57 09 59.81  & $17.35 \pm 0.07$ & $15.50 \pm 0.02$ & $15.03 \pm 0.02$ & \ldots & \ldots & $14.11 \pm 0.03$ & $14.08 \pm 0.04$ & $14.14 \pm 0.06$ &  wd \\
CM-DRA-REF58  &  16 33 44.447  &  +57 12 46.29  & $15.36 \pm 0.07$ & $15.28 \pm 0.02$ & $14.58 \pm 0.02$ & \ldots & \ldots & $13.24 \pm 0.02$ & $12.82 \pm 0.03$ & $12.75 \pm 0.03$ &  G5-G6V  \\
CM-DRA-REF65  &  16 34 22.925  &  +57 11 46.70  & $16.38 \pm 0.07$ & $16.17 \pm 0.02$ & $15.32 \pm 0.01$ & \ldots & \ldots & $13.76 \pm 0.03$ & $13.29 \pm 0.03$ & $13.33 \pm 0.03$ &  K0V  \\
CM-DRA-REF66  &  16 33 48.569  &  +57 11 40.86  & $13.38 \pm 0.07$ & $13.46 \pm 0.01$ & $12.91 \pm 0.01$ & \ldots & \ldots & $11.83 \pm 0.02$ & $11.58 \pm 0.03$ & $11.53 \pm 0.02$ &  F9V  \\
CM-DRA-REF94  &  16 34 24.625  &  +57 08 26.94  & $15.65 \pm 0.07$ & $15.33 \pm 0.01$ & $14.37 \pm 0.01$ & \ldots & \ldots & $12.59 \pm 0.02$ & $12.04 \pm 0.02$ & $11.94 \pm 0.02$ &  K0III  \\
GU-BOO  &  15 21 54.838  &  +33 56 09.25  & \ldots & $14.90 \pm 0.02$ & $13.65 \pm 0.02$ & $12.92 \pm 0.11$ & $12.09 \pm 0.07$ & $11.05 \pm 0.02$ & $10.36 \pm 0.03$ & $10.22 \pm 0.02$ &  M0/M1.5V \\
GU-BOO-REF01  &  15 22 01.145  &  +34 01 05.89  & $15.77 \pm 0.07$ & $15.56 \pm 0.02$ & $14.65 \pm 0.02$ & \ldots & \ldots & $12.98 \pm 0.03$ & $12.52 \pm 0.04$ & $12.44 \pm 0.03$ &  K0V  \\
GU-BOO-REF05  &  15 21 50.732  &  +34 00 32.70  & $16.60 \pm 0.07$ & $16.35 \pm 0.05$ & $15.40 \pm 0.02$ & \ldots & \ldots & $13.75 \pm 0.03$ & $13.19 \pm 0.04$ & $13.13 \pm 0.03$ &  K0-K2V  \\
GU-BOO-REF08  &  15 22 05.555  &  +34 00 02.18  & $14.14 \pm 0.07$ & $14.44 \pm 0.03$ & $13.85 \pm 0.03$ & \ldots & \ldots & $12.49 \pm 0.02$ & $12.13 \pm 0.03$ & $12.03 \pm 0.02$ &  F9V  \\
GU-BOO-REF25  &  15 21 58.531  &  +34 00 10.86  & $17.01 \pm 0.07$ & $16.29 \pm 0.03$ & $15.27 \pm 0.01$ & \ldots & \ldots & $13.28 \pm 0.02$ & $12.70 \pm 0.03$ & $12.57 \pm 0.02$ &  K3V  \\
GU-BOO-REF51  &  15 22 20.585  &  +33 52 14.83  & $15.02 \pm 0.07$ & $14.47 \pm 0.05$ & $13.31 \pm 0.06$ & \ldots & \ldots & $11.37 \pm 0.02$ & $10.87 \pm 0.02$ & $10.77 \pm 0.02$ &  K3V  \\
GU-BOO-REF53  &  15 21 56.870  &  +33 58 37.57  & $14.99 \pm 0.07$ & $15.24 \pm 0.03$ & $14.67 \pm 0.02$ & \ldots & \ldots & $13.51 \pm 0.02$ & $13.15 \pm 0.04$ & $13.08 \pm 0.03$ &  G0V  \\
GU-BOO-REF57  &  15 21 57.397  &  +33 55 29.10  & $15.83 \pm 0.07$ & $15.95 \pm 0.03$ & $15.32 \pm 0.01$ & \ldots & \ldots & $14.12 \pm 0.03$ & $13.79 \pm 0.03$ & $13.77 \pm 0.04$ &  G2V  \\
NSVS0103  &  13 45 34.953  &  +79 23 48.61  & \ldots & $15.06 \pm 0.23$ & $13.52 \pm 0.09$ & \ldots & \ldots & $9.69 \pm 0.02$ & $9.02 \pm 0.02$ & $8.78 \pm 0.02$ &  M2-M3V \\
NSVS0103-REF00  &  13 44 02.121  &  +79 23 49.48  & \ldots & $14.54 \pm 0.01$ & $13.76 \pm 0.01$ & \ldots & \ldots & $12.39 \pm 0.02$ & $12.01 \pm 0.02$ & $11.96 \pm 0.02$ &  G8V  \\
NSVS0103-REF01  &  13 46 21.316  &  +79 21 34.95  & \ldots & $15.00 \pm 0.02$ & $14.30 \pm 0.01$ & \ldots & \ldots & $13.02 \pm 0.02$ & $12.69 \pm 0.02$ & $12.59 \pm 0.03$ &  G5V  \\
NSVS0103-REF02  &  13 47 30.350  &  +79 23 37.52  & \ldots & $11.47 \pm 0.01$ & $10.77 \pm 0.01$ & \ldots & \ldots & $9.50 \pm 0.02$ & $9.17 \pm 0.02$ & $9.05 \pm 0.02$ &  G5V  \\
NSVS0103-REF05  &  13 45 41.357  &  +79 21 05.38  & \ldots & $15.70 \pm 0.02$ & $14.78 \pm 0.01$ & \ldots & \ldots & $13.08 \pm 0.02$ & $12.61 \pm 0.02$ & $12.49 \pm 0.02$ &  K2V  \\
NSVS0103-REF19  &  13 46 37.263  &  +79 18 47.62  & \ldots & $14.37 \pm 0.02$ & $13.66 \pm 0.01$ & \ldots & \ldots & $12.29 \pm 0.02$ & $11.91 \pm 0.02$ & $11.79 \pm 0.02$ &  G3V  \\
NSVS0103-REF23  &  13 45 20.334  &  +79 18 01.73  & \ldots & $13.20 \pm 0.01$ & $12.62 \pm 0.01$ & \ldots & \ldots & $11.53 \pm 0.02$ & $11.25 \pm 0.02$ & $11.17 \pm 0.02$ &  G0V  \\
NSVS0103-REF68  &  13 45 55.139  &  +79 23 14.00  & \ldots & $17.30 \pm 0.04$ & $15.78 \pm 0.02$ & \ldots & \ldots & $11.64 \pm 0.02$ & $11.03 \pm 0.02$ & $10.76 \pm 0.02$ &  M4.5V  \\
NSVS0103-REF96  &  13 46 58.399  &  +79 22 04.23  & \ldots & $13.52 \pm 0.02$ & $13.02 \pm 0.01$ & \ldots & \ldots & $12.00 \pm 0.02$ & $11.76 \pm 0.02$ & $11.70 \pm 0.02$ &  F6V  \\
TRES-HER0  &  16 50 20.771  &  +46 39 01.61  & \ldots & $17.23 \pm 0.05$ & $15.76 \pm 0.03$ & $14.58 \pm 0.02$ & $13.14 \pm 0.04$ & $11.77 \pm 0.02$ & $11.14 \pm 0.02$ & $10.88 \pm 0.02$ &  M2-M3 V \\
TRES-HER0-REF337  &  16 50 03.823  &  +46 43 17.02  & \ldots & $11.86 \pm 0.00$ & $11.39 \pm 0.01$ & \ldots & \ldots & $10.41 \pm 0.02$ & $10.17 \pm 0.02$ & $10.16 \pm 0.02$ &  F5V  \\
TRES-HER0-REF347  &  16 50 13.877  &  +46 41 26.68  & $14.25 \pm 0.07$ & $14.48 \pm 0.01$ & $13.88 \pm 0.01$ & \ldots & \ldots & $12.64 \pm 0.02$ & $12.30 \pm 0.02$ & $12.26 \pm 0.02$ &  G0V  \\
TRES-HER0-REF363  &  16 50 58.817  &  +46 38 49.20  & $15.68 \pm 0.07$ & $15.17 \pm 0.01$ & $14.22 \pm 0.01$ & \ldots & \ldots & $12.41 \pm 0.02$ & $11.87 \pm 0.02$ & $11.77 \pm 0.02$ &  K3V  \\
TRES-HER0-REF377  &  16 50 50.869  &  +46 36 46.17  & $14.83 \pm 0.07$ & $14.43 \pm 0.01$ & $13.56 \pm 0.01$ & \ldots & \ldots & $12.03 \pm 0.02$ & $11.65 \pm 0.02$ & $11.55 \pm 0.02$ &  K0V  \\
TRES-HER0-REF380  &  16 50 48.218  &  +46 36 32.09  & $14.48 \pm 0.07$ & $14.67 \pm 0.01$ & $14.12 \pm 0.01$ & \ldots & \ldots & $13.03 \pm 0.02$ & $12.77 \pm 0.03$ & $12.72 \pm 0.03$ &  F8III  \\
TRES-HER0-REF394  &  16 49 56.216  &  +46 33 36.03  & $12.75 \pm 0.07$ & $12.01 \pm 0.00$ & $10.93 \pm 0.01$ & \ldots & \ldots & $9.02 \pm 0.03$ & $8.51 \pm 0.02$ & $8.39 \pm 0.02$ &  K2IIIab  \\
TRES-HER0-REF823  &  16 50 44.114  &  +46 37 11.08  & $15.96 \pm 0.07$ & $16.20 \pm 0.03$ & $15.47 \pm 0.01$ & \ldots & \ldots & $13.94 \pm 0.03$ & $13.53 \pm 0.03$ & $13.44 \pm 0.05$ &  G2V  \\
YY-GEM  &  07 34 37.446  &  +31 52 10.22  & \ldots & \ldots & \ldots & \ldots & \ldots & $6.07 \pm 0.02$ & $5.42 \pm 0.02$ & $5.24 \pm 0.02$ &  M1.0Ve \\
YY-GEM-REF01  &  07 34 52.068  &  +31 52 34.14  & \ldots & \ldots & \ldots & \ldots & \ldots & $9.34 \pm 0.02$ & $8.83 \pm 0.03$ & $8.69 \pm 0.02$ &  K0IIIb  \\
YY-GEM-REF02  &  07 34 49.027  &  +31 51 42.87  & \ldots & \ldots & \ldots & \ldots & \ldots & $11.86 \pm 0.02$ & $11.56 \pm 0.02$ & $11.49 \pm 0.02$ &  G3V  \\
YY-GEM-REF03  &  07 34 26.315  &  +31 51 00.12  & \ldots & \ldots & \ldots & \ldots & \ldots & $8.27 \pm 0.02$ & $7.80 \pm 0.02$ & $7.69 \pm 0.02$ &  K0III  \\
YY-GEM-REF04  &  07 34 31.213  &  +31 51 16.65  & \ldots & \ldots & \ldots & \ldots & \ldots & $13.22 \pm 0.02$ & $12.94 \pm 0.03$ & $12.89 \pm 0.03$ &  G0V  \\
YY-GEM-REF05  &  07 34 41.849  &  +31 51 37.15  & \ldots & \ldots & \ldots & \ldots & \ldots & $15.19 \pm 0.05$ & $14.98 \pm 0.08$ & $14.78 \pm 0.10$ &  G0V  \\
YY-GEM-REF06  &  07 34 42.028  &  +31 51 58.49  & \ldots & \ldots & \ldots & \ldots & \ldots & $14.42 \pm 0.05$ & $14.31 \pm 0.04$ & $14.37 \pm 0.07$ &  G2V  \\
YY-GEM-REF07  &  07 34 31.228  &  +31 52 29.27  & \ldots & \ldots & \ldots & \ldots & \ldots & $12.41 \pm 0.05$ & $12.05 \pm 0.05$ & $11.89 \pm 0.02$ &  G8V  \\
\enddata


\tablecomments{For entries with `\ldots', no data were taken; photometric data collection is described in \S \ref{sec_photometry}.  Spectral types determined using the methodology described in \S \ref{sec_spectroscopy} using data from this table and Table \ref{tab_photometry2}.}


\end{deluxetable}





\subsection{42- and 31-inch Photometry}\label{sec_photometry}

%
%
%
%

CCD Photometry in the standard Johnson UBVRI and narrowband `Hale-Bopp' filter sets \citep{Farnham2000Icar..147..180F} was obtained for all our target stars and their astrometric reference stars, excepting YY Gem (see next paragraph) using Lowell Observatory's Hall 42-inch and 31-inch telescope located at Anderson Mesa.  Standard stars from \citet{Landolt1983AJ.....88..439L} and Farnham were used for atmospheric extinction correction.  Johnson B and V was collected for all the stars (Table \ref{tab_photometry}), as well as the narrowband filters  (Table \ref{tab_photometry2}).

Given the proximity of YY Gem to Castor (roughly 1 arcminute away, and 8 magnitudes brighter), the Kron photometer on the Hall 42-inch was used instead of the CCD cameras to obtain narrowband data for this target and its reference stars.  The Kron photometer is a single-element aperture photometer with a thermoelectrically cooled EMI6256 photomultiplier.  While slightly more time-consuming than CCD observing -- only one object can be observed at a time -- the nature of the instrument's sampling aperture allowed observations of YY Gem and its FGS reference stars to be conducted in a way to minimize and also allow subtraction of the scattered light from Castor.  CCD observations of these targets with the CCD cameras on the 42- and 31-inch telescopes were not possible due to scattering and saturation effects of of Castor on the imaging chip.  Kron photometer observations were compared to Farnham standards and extinction corrected and are found in Table \ref{tab_photometry2}.








\begin{deluxetable}{lccccccccccl}
\rotate




\tablecaption{\label{tab_photometry2}Target and reference star coordinates, photometry, and inferred spectral type: Hale-Bopp narrowband `comet' filters.}



\tabletypesize{\tiny}

\tablehead{\colhead{Program ID}
& \colhead{Bc} & \colhead{C2}& \colhead{C3} & \colhead{CN}
& \colhead{COp} & \colhead{Gc} & \colhead{NH} & \colhead{OH} & \colhead{Rc} & \colhead{UVc} & \colhead{Spectral}\\
\colhead{} & \colhead{(mag)} & \colhead{(mag)} & \colhead{(mag)} & \colhead{(mag)} & \colhead{(mag)} & \colhead{(mag)} & \colhead{(mag)} & \colhead{(mag)} &\colhead{(mag)} &\colhead{(mag)} & \colhead{Type} }

\startdata
CM-DRA  & \ldots & \ldots & \ldots & \ldots & \ldots & \ldots & \ldots & \ldots & \ldots & \ldots &  M4.5V \\
CM-DRA-REF45  & $14.46 \pm 0.01$ & $13.43 \pm 0.01$ & $16.40 \pm 0.06$ & \ldots & $14.42 \pm 0.03$ & \ldots & \ldots & \ldots & \ldots & \ldots &  G8V?  \\
CM-DRA-REF46  & $14.67 \pm 0.01$ & $14.38 \pm 0.01$ & $15.19 \pm 0.06$ & \ldots & $15.15 \pm 0.03$ & $14.21 \pm 0.02$ & \ldots & \ldots & $13.31 \pm 0.02$ & \ldots &  G2V  \\
CM-DRA-REF47  & $16.20 \pm 0.01$ & \ldots & \ldots & \ldots & $17.80 \pm 0.03$ & $16.22 \pm 0.02$ & \ldots & \ldots & $14.56 \pm 0.02$ & \ldots &  wd \\
CM-DRA-REF58  & $15.08 \pm 0.01$ & $14.65 \pm 0.01$ & $15.73 \pm 0.06$ & \ldots & $15.56 \pm 0.03$ & $14.61 \pm 0.02$ & \ldots & \ldots & $13.71 \pm 0.02$ & \ldots &  G5-G6V  \\
CM-DRA-REF65  & \ldots & $15.47 \pm 0.01$ & \ldots & \ldots & $16.34 \pm 0.03$ & $15.34 \pm 0.02$ & \ldots & \ldots & \ldots & \ldots &  K0V  \\
CM-DRA-REF66  & $13.36 \pm 0.01$ & $12.91 \pm 0.01$ & $13.77 \pm 0.06$ & \ldots & $13.57 \pm 0.03$ & $12.88 \pm 0.02$ & \ldots & \ldots & $12.13 \pm 0.02$ & \ldots &  F9V  \\
CM-DRA-REF94  & $15.13 \pm 0.01$ & $14.48 \pm 0.01$ & $15.71 \pm 0.06$ & \ldots & $15.77 \pm 0.03$ & $14.36 \pm 0.02$ & \ldots & \ldots & $13.28 \pm 0.02$ & \ldots &  K0III  \\
GU-BOO  & \ldots & \ldots & \ldots & \ldots & \ldots & \ldots & \ldots & \ldots & \ldots & \ldots &  M0/M1.5V \\
GU-BOO-REF01  & \ldots & \ldots & $15.47 \pm 0.06$ & \ldots & $15.41 \pm 0.03$ & \ldots & \ldots & \ldots & \ldots & \ldots &  K0V  \\
GU-BOO-REF05  & $15.93 \pm 0.01$ & $15.61 \pm 0.01$ & $16.33 \pm 0.06$ & \ldots & $16.12 \pm 0.03$ & $15.42 \pm 0.02$ & \ldots & \ldots & $14.36 \pm 0.02$ & \ldots &  K0-K2V  \\
GU-BOO-REF08  & $14.31 \pm 0.01$ & $13.90 \pm 0.01$ & $14.74 \pm 0.06$ & \ldots & $14.58 \pm 0.03$ & $13.83 \pm 0.02$ & \ldots & \ldots & $12.94 \pm 0.02$ & \ldots &  F9V  \\
GU-BOO-REF25  & $16.17 \pm 0.01$ & $15.62 \pm 0.01$ & \ldots & \ldots & $16.89 \pm 0.03$ & $15.20 \pm 0.02$ & \ldots & \ldots & \ldots & \ldots &  K3V  \\
GU-BOO-REF51  & $14.17 \pm 0.01$ & $13.62 \pm 0.01$ & $15.06 \pm 0.06$ & \ldots & $14.80 \pm 0.03$ & $13.36 \pm 0.02$ & \ldots & \ldots & $12.14 \pm 0.02$ & \ldots &  K3V  \\
GU-BOO-REF53  & $15.09 \pm 0.01$ & $14.70 \pm 0.01$ & $15.32 \pm 0.06$ & \ldots & $15.38 \pm 0.03$ & $14.58 \pm 0.02$ & \ldots & \ldots & $13.86 \pm 0.02$ & \ldots &  G0V  \\
GU-BOO-REF57  & $15.58 \pm 0.01$ & $15.25 \pm 0.01$ & $16.74 \pm 0.06$ & \ldots & $15.80 \pm 0.03$ & $15.20 \pm 0.02$ & \ldots & \ldots & $14.23 \pm 0.02$ & \ldots &  G2V  \\
NSVS0103  & $15.14 \pm 0.34$ & $13.89 \pm 0.09$ & $16.22 \pm 0.85$ & $16.47 \pm 0.90$ & \ldots & $13.64 \pm 0.12$ & \ldots & \ldots & \ldots & \ldots &  M2-M3V \\
NSVS0103-REF00  & $14.33 \pm 0.16$ & $13.79 \pm 0.08$ & $15.35 \pm 0.38$ & $15.12 \pm 0.24$ & \ldots & $13.54 \pm 0.11$ & \ldots & \ldots & \ldots & \ldots &  G8V  \\
NSVS0103-REF01  & $15.00 \pm 0.29$ & $14.52 \pm 0.15$ & $15.12 \pm 0.30$ & $16.05 \pm 0.59$ & \ldots & $14.27 \pm 0.20$ & \ldots & \ldots & \ldots & \ldots &  G5V  \\
NSVS0103-REF02  & $11.29 \pm 0.02$ & $10.88 \pm 0.02$ & $11.85 \pm 0.02$ & $12.48 \pm 0.03$ & \ldots & $10.75 \pm 0.02$ & $12.55 \pm 0.06$ & \ldots & \ldots & \ldots &  G5V  \\
NSVS0103-REF05  & $15.21 \pm 0.35$ & $14.85 \pm 0.21$ & \ldots & \ldots & \ldots & \ldots & \ldots & \ldots & \ldots & \ldots &  K2V  \\
NSVS0103-REF19  & $14.50 \pm 0.19$ & $13.65 \pm 0.07$ & $14.86 \pm 0.25$ & $15.03 \pm 0.24$ & \ldots & $13.70 \pm 0.13$ & $15.37 \pm 0.64$ & \ldots & \ldots & \ldots &  G3V  \\
NSVS0103-REF23  & $13.14 \pm 0.06$ & $12.67 \pm 0.03$ & $13.58 \pm 0.08$ & $13.91 \pm 0.09$ & \ldots & $12.57 \pm 0.05$ & $14.09 \pm 0.20$ & \ldots & \ldots & \ldots &  G0V  \\
NSVS0103-REF68  & \ldots & $15.47 \pm 0.36$ & \ldots & \ldots & \ldots & \ldots & \ldots & \ldots & \ldots & \ldots &  M4.5V  \\
NSVS0103-REF96  & $13.32 \pm 0.07$ & $13.03 \pm 0.04$ & $13.79 \pm 0.09$ & $14.04 \pm 0.11$ & \ldots & $12.98 \pm 0.07$ & \ldots & \ldots & \ldots & \ldots &  F6V  \\
TRES-HER0  & \ldots & \ldots & \ldots & \ldots & \ldots & \ldots & \ldots & \ldots & \ldots & \ldots &  M2-M3 V \\
TRES-HER0-REF337  & \ldots & \ldots & \ldots & \ldots & \ldots & \ldots & \ldots & \ldots & \ldots & \ldots &  F5V  \\
TRES-HER0-REF347  & $14.31 \pm 0.01$ & $13.92 \pm 0.01$ & $14.69 \pm 0.06$ & \ldots & $14.54 \pm 0.03$ & $13.84 \pm 0.02$ & \ldots & \ldots & $13.04 \pm 0.02$ & \ldots &  G0V  \\
TRES-HER0-REF363  & $14.98 \pm 0.01$ & $14.59 \pm 0.01$ & $15.84 \pm 0.06$ & \ldots & $15.61 \pm 0.03$ & $14.25 \pm 0.02$ & \ldots & \ldots & $13.12 \pm 0.02$ & \ldots &  K3V  \\
TRES-HER0-REF377  & $14.21 \pm 0.01$ & $13.74 \pm 0.01$ & $14.86 \pm 0.06$ & \ldots & $14.70 \pm 0.03$ & $13.60 \pm 0.02$ & \ldots & \ldots & $12.61 \pm 0.02$ & \ldots &  K0V  \\
TRES-HER0-REF380  & $14.51 \pm 0.01$ & $14.13 \pm 0.01$ & $14.97 \pm 0.06$ & \ldots & $14.76 \pm 0.03$ & $14.02 \pm 0.02$ & \ldots & \ldots & $13.33 \pm 0.02$ & \ldots &  F8III  \\
TRES-HER0-REF394  & $11.86 \pm 0.01$ & $11.12 \pm 0.01$ & $12.73 \pm 0.06$ & \ldots & $12.31 \pm 0.03$ & $11.01 \pm 0.02$ & \ldots & \ldots & $9.87 \pm 0.02$ & \ldots &  K2IIIab  \\
TRES-HER0-REF823  & $15.98 \pm 0.01$ & $15.42 \pm 0.01$ & $16.28 \pm 0.06$ & \ldots & $16.29 \pm 0.03$ & $15.43 \pm 0.02$ & \ldots & \ldots & $14.53 \pm 0.02$ & \ldots &  G2V  \\
YY-GEM  & $10.33 \pm 0.02$ & $9.65 \pm 0.01$ & $11.40 \pm 0.02$ & $12.08 \pm 0.03$ & \ldots & $9.24 \pm 0.02$ & $12.62 \pm 0.05$ & $13.37 \pm 0.23$ & \ldots & $12.38 \pm 0.04$ &  M1.0Ve \\
YY-GEM-REF01  & $12.16 \pm 0.03$ & $11.44 \pm 0.02$ & $12.97 \pm 0.04$ & $14.05 \pm 0.07$ & \ldots & $11.27 \pm 0.02$ & $14.19 \pm 0.11$ & \ldots & \ldots & $14.30 \pm 0.10$ &  K0IIIb  \\
YY-GEM-REF02  & $13.61 \pm 0.07$ & $13.07 \pm 0.04$ & $13.91 \pm 0.08$ & $15.16 \pm 0.23$ & \ldots & $13.05 \pm 0.07$ & $14.89 \pm 0.25$ & $15.15 \pm 0.82$ & \ldots & $14.80 \pm 0.17$ &  G3V  \\
YY-GEM-REF03  & $10.77 \pm 0.02$ & $10.12 \pm 0.01$ & $11.49 \pm 0.02$ & $12.50 \pm 0.03$ & \ldots & $10.03 \pm 0.02$ & $12.80 \pm 0.05$ & \ldots & \ldots & $12.49 \pm 0.03$ &  K0III  \\
YY-GEM-REF04  & $14.42 \pm 0.13$ & \ldots & $15.00 \pm 0.21$ & $14.83 \pm 0.14$ & \ldots & $14.16 \pm 0.15$ & $15.51 \pm 0.46$ & \ldots & \ldots & $15.38 \pm 0.28$ &  G0V  \\
YY-GEM-REF05  & \ldots & \ldots & \ldots & \ldots & \ldots & \ldots & \ldots & \ldots & \ldots & \ldots &  G0V  \\
YY-GEM-REF06  & $15.16 \pm 0.30$ & $15.03 \pm 0.23$ & \ldots & \ldots & \ldots & $15.38 \pm 0.56$ & \ldots & $15.29 \pm 1.07$ & \ldots & \ldots &  G2V  \\
YY-GEM-REF07  & $13.73 \pm 0.08$ & $13.03 \pm 0.04$ & $14.06 \pm 0.10$ & $14.19 \pm 0.10$ & \ldots & $12.99 \pm 0.06$ & $14.60 \pm 0.18$ & \ldots & \ldots & $14.59 \pm 0.14$ &  G8V  \\
\enddata


\tablecomments{For entries with `\ldots', no data were taken; photometric data collection is described in \S \ref{sec_photometry}. For details on the Hale-Bopp narrowband `comet' filters, see  \citet{Farnham2000Icar..147..180F}.Spectral types determined using the methodology described in \S \ref{sec_spectroscopy} using data from this table and Table \ref{tab_photometry}.}


\end{deluxetable}


\clearpage







\begin{deluxetable}{lcccccccccccc}
\rotate



\tablecaption{\label{tab_refStarsDist}Estimated and measured parameters for reference stars.}


\tablehead{
\colhead{Program ID} &
\colhead{Ang. Size} &
\colhead{$F_{\rm BOL}$ $^{\rm a}$}&
\colhead{$A_V$} &
\colhead{Sp. Type} &
\colhead{$R$ $^{\rm b}$} &
\colhead{$d_{\rm R}$} &
\colhead{$M_V$} &
\colhead{$BC$} &
\colhead{$d_{\rm MV}$ $^{\rm c}$} &
\colhead{$d_{\rm final}$} &
\colhead{$\pi_{\rm final}$} &
\colhead{$\pi_{Gaia}$}\\
\colhead{} &
\colhead{(mas)} &
\colhead{} &
\colhead{(mag)} &
\colhead{} &
\colhead{$(R_\odot)$} &
\colhead{(pc)} &
\colhead{(mag)} &
\colhead{(mag)} &
\colhead{(pc)} &
\colhead{(pc)} &
\colhead{(mas)} &
\colhead{(mas)}
}
\startdata
CM-DRA-REF45 & $0.0295 \pm 0.0021$ & $23.30 \pm 0.84$ & $0.75 \pm 0.02$ & G8V & 0.93 & $294 \pm 36$ & 5.5 & -0.4 & $220 \pm 44$ & $264 \pm 52$ & $3.786 \pm 0.746$ & $0.590 \pm 0.016$ \\
CM-DRA-REF46 & $0.0131 \pm 0.0009$ & $5.48 \pm 0.10$ & $0.00 \pm 0.02$ & G2V & 1.25 & $886 \pm 108$ & 4.7 & -0.2 & $847 \pm 169$ & $874 \pm 27$ & $1.144 \pm 0.036$ & $1.946 \pm 0.204$ \\
CM-DRA-REF58 & $0.0121 \pm 0.0009$ & $4.97 \pm 0.12$ & $0.18 \pm 0.02$ & G5V & 0.99 & $757 \pm 93$ & 5.1 & -0.2 & $683 \pm 137$ & $734 \pm 52$ & $1.363 \pm 0.096$ & $1.203 \pm 0.023$ \\
CM-DRA-REF65 & $0.0097 \pm 0.0007$ & $2.55 \pm 0.06$ & $0.12 \pm 0.02$ & K0V & 0.79 & $763 \pm 93$ & 5.9 & -0.3 & $711 \pm 142$ & $747 \pm 36$ & $1.338 \pm 0.065$ & $1.352 \pm 0.032$ \\
CM-DRA-REF66 & $0.0217 \pm 0.0015$ & $19.20 \pm 0.35$ & $0.04 \pm 0.02$ & F9V & 1.43 & $614 \pm 75$ & 4.2 & -0.2 & $552 \pm 110$ & $594 \pm 43$ & $1.683 \pm 0.123$ & $1.034 \pm 0.029$ \\
CM-DRA-REF94 & $0.0214 \pm 0.0015$ & $6.66 \pm 0.12$ & $0.07 \pm 0.02$ & K0III & 13.60 & $5905 \pm 723$ & 0.7 & -0.5 & $5381 \pm 1076$ & $5742 \pm 370$ & $0.174 \pm 0.011$ & $0.170 \pm 0.018$ \\
GU-BOO-REF01 & $0.0143 \pm 0.0010$ & $5.57 \pm 0.16$ & $0.29 \pm 0.02$ & K0V & 0.79 & $516 \pm 63$ & 5.9 & -0.3 & $444 \pm 89$ & $492 \pm 51$ & $2.032 \pm 0.210$ & $1.971 \pm 0.240$ \\
GU-BOO-REF05 & $0.0101 \pm 0.0007$ & $2.77 \pm 0.08$ & $0.28 \pm 0.02$ & K0V & 0.79 & $731 \pm 90$ & 5.9 & -0.3 & $635 \pm 127$ & $699 \pm 68$ & $1.430 \pm 0.138$ & $1.435 \pm 0.026$ \\
GU-BOO-REF08 & $0.0169 \pm 0.0012$ & $11.70 \pm 0.34$ & $0.41 \pm 0.02$ & F9V & 1.43 & $788 \pm 97$ & 4.2 & -0.2 & $596 \pm 119$ & $712 \pm 136$ & $1.405 \pm 0.268$ & $2.403 \pm 0.041$ \\
GU-BOO-REF25 & $0.0154 \pm 0.0011$ & $3.79 \pm 0.09$ & $0.27 \pm 0.02$ & K3V & 0.78 & $469 \pm 58$ & 6.7 & -0.5 & $412 \pm 82$ & $451 \pm 40$ & $2.220 \pm 0.197$ & $1.744 \pm 0.023$ \\
GU-BOO-REF51 & $0.0354 \pm 0.0025$ & $20.00 \pm 0.35$ & $0.12 \pm 0.02$ & K3V & 0.78 & $204 \pm 25$ & 6.7 & -0.5 & $192 \pm 38$ & $201 \pm 9$ & $4.988 \pm 0.212$ & $5.119 \pm 0.022$ \\
GU-BOO-REF53 & $0.0106 \pm 0.0008$ & $4.18 \pm 0.09$ & $0.12 \pm 0.02$ & G0V & 1.28 & $1122 \pm 137$ & 4.4 & -0.2 & $1046 \pm 209$ & $1099 \pm 54$ & $0.910 \pm 0.044$ & $1.065 \pm 0.020$ \\
GU-BOO-REF57 & $0.0084 \pm 0.0006$ & $2.26 \pm 0.04$ & $0.00 \pm 0.02$ & G2V & 1.25 & $1376 \pm 168$ & 4.7 & -0.2 & $1319 \pm 264$ & $1360 \pm 41$ & $0.736 \pm 0.022$ & $0.685 \pm 0.025$ \\
NSVS0103-REF00 & $0.0187 \pm 0.0013$ & $9.34 \pm 0.17$ & $0.02 \pm 0.02$ & G8V & 0.93 & $464 \pm 57$ & 5.5 & -0.4 & $488 \pm 98$ & $470 \pm 17$ & $2.129 \pm 0.078$ & $2.048 \pm 0.015$ \\
NSVS0103-REF01 & $0.0132 \pm 0.0009$ & $5.94 \pm 0.12$ & $0.12 \pm 0.02$ & G5V & 0.99 & $694 \pm 85$ & 5.1 & -0.2 & $643 \pm 129$ & $678 \pm 35$ & $1.474 \pm 0.077$ & $1.386 \pm 0.016$ \\
NSVS0103-REF02 & $0.0669 \pm 0.0047$ & $153.00 \pm 2.68$ & $0.12 \pm 0.02$ & G5V & 0.99 & $137 \pm 17$ & 5.1 & -0.2 & $127 \pm 25$ & $134 \pm 7$ & $7.473 \pm 0.395$ & $6.798 \pm 0.030$ \\
NSVS0103-REF05 & $0.0141 \pm 0.0010$ & $4.17 \pm 0.08$ & $0.00 \pm 0.02$ & K2V & 0.71 & $469 \pm 57$ & 6.4 & -0.4 & $491 \pm 98$ & $474 \pm 16$ & $2.108 \pm 0.071$ & $2.260 \pm 0.018$ \\
NSVS0103-REF19 & $0.0192 \pm 0.0014$ & $12.20 \pm 0.26$ & $0.24 \pm 0.02$ & G3V & 1.14 & $553 \pm 68$ & 4.8 & -0.2 & $479 \pm 96$ & $528 \pm 52$ & $1.893 \pm 0.186$ & $0.831 \pm 0.014$ \\
NSVS0103-REF23 & $0.0262 \pm 0.0019$ & $25.60 \pm 0.45$ & $0.07 \pm 0.02$ & G0V & 1.28 & $454 \pm 56$ & 4.4 & -0.2 & $432 \pm 86$ & $447 \pm 16$ & $2.235 \pm 0.079$ & $1.780 \pm 0.022$ \\
NSVS0103-REF68 & $0.0553 \pm 0.0039$ & $11.00 \pm 0.17$ & $0.00 \pm 0.03$ & M4V & 0.19 & $32 \pm 4$ & 11.5 & -2.5 & $74 \pm 15$ & $35 \pm 29$ & $28.614 \pm 23.947$ & $16.572 \pm 0.048$ \\
NSVS0103-REF96 & $0.0194 \pm 0.0014$ & $18.50 \pm 0.34$ & $0.13 \pm 0.02$ & F6V & 1.70 & $815 \pm 100$ & 3.7 & -0.1 & $682 \pm 136$ & $769 \pm 94$ & $1.301 \pm 0.159$ & $0.876 \pm 0.026$ \\
TRES-HER0-REF337 & $0.0398 \pm 0.0028$ & $82.00 \pm 1.45$ & $0.13 \pm 0.01$ & F5V & 1.83 & $428 \pm 52$ & 3.5 & -0.1 & $348 \pm 70$ & $399 \pm 56$ & $2.507 \pm 0.352$ & $1.863 \pm 0.027$ \\
TRES-HER0-REF347 & $0.0159 \pm 0.0011$ & $9.47 \pm 0.20$ & $0.22 \pm 0.02$ & G0V & 1.28 & $748 \pm 92$ & 4.4 & -0.2 & $662 \pm 132$ & $720 \pm 61$ & $1.389 \pm 0.118$ & $0.761 \pm 0.016$ \\
TRES-HER0-REF363 & $0.0219 \pm 0.0015$ & $7.67 \pm 0.12$ & $0.00 \pm 0.02$ & K3V & 0.78 & $330 \pm 40$ & 6.7 & -0.5 & $328 \pm 66$ & $329 \pm 2$ & $3.037 \pm 0.014$ & $3.053 \pm 0.023$ \\
TRES-HER0-REF377 & $0.0216 \pm 0.0015$ & $12.70 \pm 0.23$ & $0.14 \pm 0.02$ & K0V & 0.79 & $342 \pm 42$ & 5.9 & -0.3 & $316 \pm 63$ & $334 \pm 18$ & $2.995 \pm 0.166$ & $2.550 \pm 0.016$ \\
TRES-HER0-REF380 & $0.0125 \pm 0.0009$ & $6.26 \pm 0.10$ & $0.00 \pm 0.02$ & F8III & 5.25 & $3901 \pm 477$ & 1.2 & -0.3 & $4120 \pm 824$ & $3956 \pm 155$ & $0.253 \pm 0.010$ & $0.859 \pm 0.021$ \\
TRES-HER0-REF394 & $0.1200 \pm 0.0084$ & $173.00 \pm 2.37$ & $0.00 \pm 0.01$ & K2III & 19.00 & $1471 \pm 180$ & 0.5 & -0.6 & $1260 \pm 252$ & $1400 \pm 150$ & $0.714 \pm 0.076$ & $0.890 \pm 0.022$ \\
TRES-HER0-REF823 & $0.0090 \pm 0.0006$ & $2.54 \pm 0.08$ & $0.32 \pm 0.02$ & G2V & 1.25 & $1296 \pm 159$ & 4.7 & -0.2 & $1073 \pm 215$ & $1217 \pm 158$ & $0.822 \pm 0.107$ & $0.190 \pm 0.032$ \\
YY-GEM-REF01 & $0.0984 \pm 0.0070$ & $141.00 \pm 2.96$ & $0.32 \pm 0.02$ & K0III & 13.60 & $1284 \pm 157$ & 0.7 & -0.5 & $1042 \pm 208$ & $1196 \pm 171$ & $0.836 \pm 0.120$ & $0.883 \pm 0.040$ \\
YY-GEM-REF02 & $0.0223 \pm 0.0016$ & $16.30 \pm 0.29$ & $0.00 \pm 0.03$ & G3V & 1.14 & $476 \pm 58$ & 4.8 & -0.2 & $463 \pm 93$ & $472 \pm 9$ & $2.119 \pm 0.042$ & $2.444 \pm 0.036$ \\
YY-GEM-REF03 & $0.1540 \pm 0.0109$ & $346.00 \pm 5.65$ & $0.00 \pm 0.02$ & K0III & 13.60 & $821 \pm 101$ & 0.7 & -0.5 & $772 \pm 154$ & $806 \pm 34$ & $1.240 \pm 0.053$ & $1.079 \pm 0.044$ \\
YY-GEM-REF04 & $0.0120 \pm 0.0009$ & $5.36 \pm 0.14$ & $0.00 \pm 0.08$ & G0V & 1.28 & $991 \pm 122$ & 4.4 & -0.2 & $974 \pm 195$ & $987 \pm 12$ & $1.014 \pm 0.012$ & $0.704 \pm 0.032$ \\
YY-GEM-REF05 & $0.0048 \pm 0.0005$ & $0.87 \pm 0.12$ & $0.08 \pm 0.57$ & G0V & 1.28 & $2468 \pm 345$ & 4.4 & -0.2 & $2334 \pm 467$ & $2421 \pm 95$ & $0.413 \pm 0.016$ & $0.816 \pm 0.084$ \\
YY-GEM-REF06 & $0.0064 \pm 0.0005$ & $1.30 \pm 0.08$ & $0.00 \pm 0.32$ & G2V & 1.25 & $1816 \pm 229$ & 4.7 & -0.2 & $1739 \pm 348$ & $1792 \pm 54$ & $0.558 \pm 0.017$ & $ \ldots $ \\
YY-GEM-REF07 & $0.0200 \pm 0.0014$ & $10.60 \pm 0.23$ & $0.00 \pm 0.06$ & G8V & 0.93 & $433 \pm 53$ & 5.5 & -0.4 & $462 \pm 92$ & $441 \pm 20$ & $2.270 \pm 0.104$ & $2.040 \pm 0.031$ \\
\enddata
\tablecomments{Distances estimated from two methodologies: bolometric-flux estimated angular size vs. linear size estimates ($d_{\rm R}$), and bolometric-corrected estimates of $M_{\rm V}$.  For more information, see Sections \ref{sec_sedFit} and \ref{sec_inferred_distances}.}

\tablenotetext{a}{Units of $F_{\rm BOL}$ are $10^{-11}$ erg cm$^{-2}$ s$^{-1}$.}
\tablenotetext{b}{A 10\% radius error is expected from linear radii inferred from spectral type.}
\tablenotetext{c}{A 20\% distance error is expected from distances inferred from $M_V$.}

\end{deluxetable}




\section{Analysis} \label{sec_analysis}


\subsection{Spectral Energy Distribution}\label{sec_sedFit}

We performed SED fits for every target and calibrator star.  These fits take photometry described in \S \ref{sec_photometry} as input values, and subsequently fit template spectra to them. The choice of template spectra from \citet{Pickles1998PASP..110..863P} is determined by the WIYN Hydra spectra described in \S \ref{sec_spectroscopy}. Each template is adjusted to account for overall flux level, wavelength-dependent reddening, and expected angular size. Reddening corrections are based upon the empirical reddening determination described by \citet{Cardelli1989ApJ...345..245C}, which differs little from van de Hulst's
theoretical reddening curve number 15 \citep{Johnson1968nim..book..167J,Dyck1996AJ....111.1705D}.
We thus obtain estimates for every single star for regarding its bolometric flux ($F_{BOL}$) and interstellar reddening ($A_V$) along the line of sight. Based on the effective temperature values that are given for each of the \citet{Pickles1998PASP..110..863P} templates, we also calculate stellar angular size ($\theta$). The results of the fitting are given in Table \ref{tab_refStarsDist}.




\subsection{Binary SED data from {\tt binarySED}} \label{sec_binarysed}



An extension of the {\tt sedFit} code discussed in \S \ref{sec_sedFit}, {\tt binarySED}, was developed for fitting of pairs of SED curves that make up a binary's observed, blended SED.  Just as the {\tt sedFit} code can search for the individual best fit template for a given input set of photometry, {\tt binarySED} will search using pairs of templates.  Input data for {\tt binarySED} is also broad- to narrow-band photometry, as well as {\it a priori} primary-secondary flux ratio estimates at defined bandpasses.  The best-fit spectral templates for the two components of each binary are presented in Table \ref{tab_binaryResults}.

The wavelength-dependent flux ratio estimates for the binaries are given in Table \ref{tab_ratio_data}.  For GU Boo, there is a particularly rich data set from \citet{Lopez-Morales2005ApJ...631.1120L}, covering multiple narrow bands from 532 to 714 nm.  Other stars, such as CM Dra and NSVS0103, only have estimates for the $V$-, $R$-, and $I$-bands.




\begin{deluxetable*}{lcc|ccl|cc|}




\tablecaption{Binary SEDfit results from this study on the systems of interest, along with literature values: parallax/distance, and primary/secondary spectral types.\label{tab_binaryResults}}


\tablehead{\colhead{} & \colhead{} &\colhead{} &\multicolumn{3}{|c|}{Literature}& \multicolumn{2}{|c|}{{\tt binarySED}}\\
\colhead{Target} & \colhead{$\pi$} & \colhead{$d$} & \colhead{Primary}& \colhead{Secondary} & \colhead{Reference} & \colhead{Primary}& \colhead{Secondary}\\
\colhead{} & \colhead{(mas)} & \colhead{(pc)} & \colhead{SpType}& \colhead{SpType} & \colhead{} & \colhead{SpType}& \colhead{SpType} }  
\startdata
TrES-Her0 & $6.06 \pm 0.47$ & $165.0 \pm 12.8$ & K9.5 & K9.5 & \citet{Creevey2005ApJ...625L.127C} & M3 & M3.5 \\
CM Dra & $68.48 \pm 0.43$ & $14.60 \pm 0.09$ & M5 & M5.5 & \citet{Metcalfe1996ApJ...456..356M} & M4 & M4 \\
GU Boo & $3.19 \pm 0.60$ & $313.5 \pm 59.0$ & K7 & K7 & \citet{Lopez-Morales2005ApJ...631.1120L} & K8.5 & M0 \\
NSVS0103 & $15.29 \pm 0.58$ & $65.40 \pm 2.48$ & K8.5 & K9.5 & \citet{Lopez-Morales2006astro.ph.10225L} & M1.5 & M3.5 \\
YY Gem & $67.22 \pm 0.42$ & $14.88 \pm 0.09$ & K7 & K7 & \citet{Torres2002ApJ...567.1140T} & M0.5 & M0.5 \\
\enddata

\tablecomments{For more information, see \S \ref{sec_binarysed}.}



\end{deluxetable*}




\begin{deluxetable*}{lcccl}




\tablecaption{Primary-secondary photometry ratio data used as input to {\tt binarySED}.\label{tab_ratio_data}}


\tablehead{\colhead{Target} & \colhead{Wavelength} & \colhead{Bandpass} & \colhead{Ratio}& \colhead{Reference}\\
\colhead{} & \colhead{(nm)} & \colhead{(nm)} & \colhead{}& \colhead{}}  
\startdata
GU Boo & 532.5 & 10 & $0.75 \pm  0.1$ & \citet{Lopez-Morales2005ApJ...631.1120L} (Figure 3) \\
GU Boo & 552.5 & 10 & $0.83 \pm  0.1$ & \citet{Lopez-Morales2005ApJ...631.1120L} (Figure 3) \\
GU Boo & 562.5 & 10 & $0.88 \pm  0.125$ & \citet{Lopez-Morales2005ApJ...631.1120L} (Figure 3) \\
GU Boo & 572.5 & 10 & $0.77 \pm  0.15$ & \citet{Lopez-Morales2005ApJ...631.1120L} (Figure 3) \\
GU Boo & 584 & 10 & $0.875 \pm  0.15$ & \citet{Lopez-Morales2005ApJ...631.1120L} (Figure 3) \\
GU Boo & 585 & 10 & $0.79 \pm  0.15$ & \citet{Lopez-Morales2005ApJ...631.1120L} (Figure 3) \\
GU Boo & 593 & 10 & $0.77 \pm  0.12$ & \citet{Lopez-Morales2005ApJ...631.1120L} (Figure 3) \\
GU Boo & 595 & 10 & $0.77 \pm  0.14$ & \citet{Lopez-Morales2005ApJ...631.1120L} (Figure 3) \\
GU Boo & 606 & 10 & $0.82 \pm  0.2$ & \citet{Lopez-Morales2005ApJ...631.1120L} (Figure 3) \\
GU Boo & 612 & 10 & $0.8 \pm  0.15$ & \citet{Lopez-Morales2005ApJ...631.1120L} (Figure 3) \\
GU Boo & 620 & 10 & $0.925 \pm  0.1$ & \citet{Lopez-Morales2005ApJ...631.1120L} (Figure 3) \\
GU Boo & 626 & 10 & $0.97 \pm  0.125$ & \citet{Lopez-Morales2005ApJ...631.1120L} (Figure 3) \\
GU Boo & 647.5 & 10 & $0.92 \pm  0.125$ & \citet{Lopez-Morales2005ApJ...631.1120L} (Figure 3) \\
GU Boo & 662 & 10 & $0.87 \pm  0.175$ & \citet{Lopez-Morales2005ApJ...631.1120L} (Figure 3) \\
GU Boo & 671 & 10 & $0.79 \pm  0.175$ & \citet{Lopez-Morales2005ApJ...631.1120L} (Figure 3) \\
GU Boo & 705 & 10 & $0.86 \pm  0.2$ & \citet{Lopez-Morales2005ApJ...631.1120L} (Figure 3) \\
GU Boo & 714 & 10 & $0.86 \pm  0.19$ & \citet{Lopez-Morales2005ApJ...631.1120L} (Figure 3) \\
CM Dra & 555 & 89 & $0.871 \pm  0.125$ & \citet{Chabrier1995ApJ...451L..29C} (Table 1) \\
CM Dra & 693 & 210 & $0.8721 \pm  0.032$ & \citet{Morales2009ApJ...691.1400M} (Table 5) \\
CM Dra & 879 & 171 & $0.8782 \pm  0.033$ & \citet{Morales2009ApJ...691.1400M} (Table 5) \\
NSVS0103 & 555 & 89 & $0.95 \pm  0.05$ & \citet{Lopez-Morales2006astro.ph.10225L}  \\
NSVS0103 & 693 & 210 & $1 \pm  0.05$ & \citet{Lopez-Morales2006astro.ph.10225L}  \\
NSVS0103 & 879 & 171 & $1 \pm  0.05$ & \citet{Lopez-Morales2006astro.ph.10225L}  \\
TRES-HER0 & 555 & 89 & $0.912 \pm  0.046$ & \citet{Creevey2005ApJ...625L.127C} (Table 3) \\
TRES-HER0 & 693 & 210 & $0.92 \pm  0.06$ & \citet{Creevey2005ApJ...625L.127C} (Table 3) \\
TRES-HER0 & 879 & 171 & $0.944 \pm  0.08$ & \citet{Creevey2005ApJ...625L.127C} (Table 3) \\
TRES-HER0 & 1250 & 250 & $1 \pm  0.1$ & \citet{Creevey2005ApJ...625L.127C} (Table 3) \\
TRES-HER0 & 1650 & 300 & $1 \pm  0.1$ & \citet{Creevey2005ApJ...625L.127C} (Table 3) \\
TRES-HER0 & 2200 & 400 & $1 \pm  0.1$ & \citet{Creevey2005ApJ...625L.127C} (Table 3) \\
YY GEM & 555 & 89 & $1 \pm  0.01$ & \citet{Torres2002ApJ...567.1140T} (Table 4), \citet{Butler2015MNRAS.446.4205B} \\
YY GEM & 693 & 210 & $1 \pm  0.01$ & \citet{Torres2002ApJ...567.1140T} (Table 4), \citet{Butler2015MNRAS.446.4205B} \\
YY GEM & 879 & 171 & $1 \pm  0.01$ & \citet{Torres2002ApJ...567.1140T} (Table 4), \citet{Butler2015MNRAS.446.4205B} \\
\enddata

\tablecomments{For more information on how this ratio data was used in binary sedFits of individual systems, see \S\ref{sec_binarysed}.}



\end{deluxetable*}

The output from {\tt binarySED} is, separately for both the primary and secondary star, similar to the {\tt sedFit} output.  This includes an estimate of interstellar extinction for the binary pair ($A_{\rm V}$), and for each component, a measurement of the bolometric flux ($F_{\rm BOL}$). Based upon the effective temperature defined for the template that fits best for each stellar component, the apparent angular size is estimated ($\theta_{\rm EST}$).  These results are presented in Table \ref{tab_binarySEDdata}.




\begin{deluxetable*}{lcc|cc|cc|}




\tablecaption{Primary and secondary fit parameters from {\tt binarySED}.\label{tab_binarySEDdata}}


\tablehead{\colhead{} & \colhead{} & \colhead{} & \multicolumn{2}{|c|}{Primary}& \multicolumn{2}{|c|}{Secondary}\\
\colhead{Target} & \colhead{$\chi^2$} & \colhead{System $F_{\rm BOL}$} & \colhead{$F_{\rm BOL}$}& \colhead{$\theta$}& \colhead{$F_{\rm BOL}$}& \colhead{$\theta$}}  
\startdata
TrES-Her0 & 12.55 & $(8.97 \pm 0.10) \times 10^{-11}$ & $(4.36 \pm 0.07) \times 10^{-11}$ & $0.031 \pm 0.002$ & $(4.61 \pm 0.07) \times 10^{-11}$ & $0.034 \pm 0.002$ \\
CM Dra & 22.36 & $(1.62 \pm 0.02) \times 10^{-09}$ & $(8.62 \pm 0.12) \times 10^{-10}$ & $0.155 \pm 0.011$ & $(7.54 \pm 0.12) \times 10^{-10}$ & $0.145 \pm 0.010$ \\
GU Boo & 13.74 & $(2.21 \pm 0.01) \times 10^{-10}$ & $(1.12 \pm 0.01) \times 10^{-10}$ & $0.038 \pm 0.002$ & $(1.09 \pm 0.01) \times 10^{-10}$ & $0.040 \pm 0.002$ \\
NSVS0103 & 2.59 & $(6.36 \pm 0.32) \times 10^{-10}$ & $(3.57 \pm 0.28) \times 10^{-10}$ & $0.091 \pm 0.007$ & $(2.79 \pm 0.14) \times 10^{-10}$ & $0.085 \pm 0.007$ \\
YY Gem & 16.81 & $(1.88 \pm 0.01) \times 10^{-08}$ & $(1.06 \pm 0.01) \times 10^{-08}$ & $0.411 \pm 0.029$ & $(9.52 \pm 0.05) \times 10^{-09}$ & $0.390 \pm 0.027$ \\
\enddata
\tablecomments{Units of $F_{\rm BOL}$ are erg cm$^{-2}$ s$^{-1}$ and $\theta$ are mas; estimates from {\tt binarySED} include overall system bolometric flux (left column), and primary/secondary star fluxes and angular sizes (middle and right columns, respectively). For more information, see \S\ref{sec_binarysed}.}




\end{deluxetable*}



\subsection{Inferred Reference Star Distances}\label{sec_inferred_distances}



Following the SED fitting described in \S \ref{sec_sedFit}, distances to the parallactic reference stars were estimated in two ways and are presented in Table \ref{tab_refStarsDist}.  These two ways we interpreted the {\tt sedFit} data are as follows:
First, we compared the estimated angular size to linear size expected from the spectral type and compute the distance ($d_R$), since $d \propto R / \theta$.  Second, we compared the \S \ref{sec_sedFit} estimate of bolometric flux to the intrinsic bolometric flux expected from spectral type with the usual relationship, $m_{\rm V} - M_{\rm V} = 5 \log d + A_V$.
From the {\tt sedFit} measured bolometric flux $F_{\rm BOL}$, the bolometric magnitude $m_{\rm BOL}$ can be computed with the standard zeropoint of $2.48 \times 10^{-5}$ erg cm$^{-2}$ s$^{-1}$ for $m_{\rm BOL}=0$ \citep{Cox2000asqu.book.....C}.  For a given spectral type, \citet{Cox2000asqu.book.....C} can also be referenced for values of $M_{\rm V}$ and $BC$; with $BC$, the $m_{\rm BOL}$ can be converted to $m_V$.  Thus, all the elements for a flux-based estimate of the distance ($d_{MV}$) are obtained.

\begin{figure}
\plotone{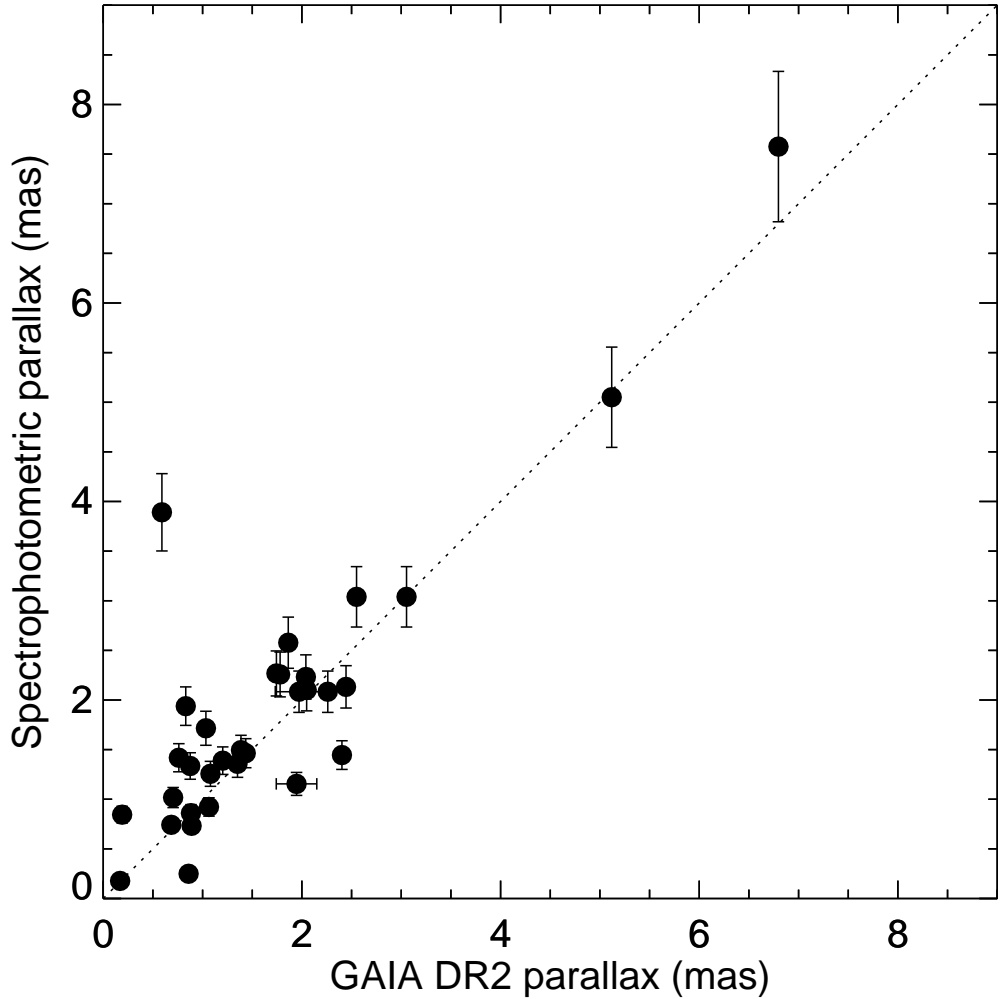}
\caption{\label{fig_par_gaia_phot} Comparison of the spectrophotometric distances of the reference stars with {\it Gaia} DR2 parallaxes. The outlier / most discrepant result is CM-Dra-REF45; the reasons for the discrepancy are not known. For more details, see Section \ref{sec_inferred_distances}.}
\end{figure}

In Table~\ref{tab_refStarsDist} we list the spectral type, stellar radius, distance estimate $d_R$, $M_V$, $BC$, and distance estimate $d_{MV}$.  The last column in Table presents the final adopted distances determined from an average of the two methods.  Figure~\ref{fig_par_gaia_phot} shows a comparison of these reference star distances with parallaxes from the {\it Gaia} DR2 catalog \citep{Gaia2018A&A...616A...1G}.



\subsection{Trigonometric Parallaxes of FGS Target Stars}\label{sec_trig_parallax}

An astrometric model was fit to the FGS positions measured in \S \ref{sec_HSTobservations} using the least squares estimation program {\tt GaussFit} \citep{Jeffrys1988CeMec..41...39J} following the method described by \citet{Bond2013ApJ...765L..12B} and \citet{Benedict2017PASP..129a2001B}.  The model consists of a four parameter plate solution for each FGS epoch that accounts for the translation, rotation, and scale differences in the field of view during each observation.  It also includes a lateral color correction \citep{Benedict1999AJ....118.1086B,Nelan2013ApJ...773L..26N}.  The model accounts for the proper motion and parallax of each star in the field of view. An expanded six parameter plate solution, which accommodates the additional adjustment of the relative scales along the x and y axis independently, did not change and/or improve our solutions by statistically significant amounts.

As input to {\tt GaussFit}, we computed parallaxes for the reference stars from the distance estimates listed in Table~\ref{tab_refStarsDist} and assumed an uncertainty of 20\%, which was consistent with the general spread between the two estimation techniques.  The initial proper motions for the reference stars were collected from several catalogs including UCAC4 \citep{Zacharias2013AJ....145...44Z}, UCAC5 \citep{Zacharias2017AJ....153..166Z}, URAT \citep{Zacharias2015AJ....150..101Z}, and PPMXL \citep{Roeser2010AJ....139.2440R}.  The UCAC5 presents an improved astrometric solution by combining the existing UCAC data with the first {\it Gaia} data release \citep{Gaia2016A&A...595A...1G}.  The high proper motion star CM-DRA-REF47 was not included in the UCAC4, UCAC5, and URAT catalogs, so we adopted the proper motion for that star from \citet{Lepine2005AJ....129.1483L}.  We fit the FGS astrometric data using each set of initial proper motions.  The {\tt GaussFit} code varies the input parallaxes and proper motions of the reference stars within their uncertainties to solve for the best astrometric grid.  The output proper motions showed a strong dependence on the input values, representing an absolute shift between different catalogs, as was discussed by \citet{Bond2013ApJ...765L..12B}.  However, the parallaxes of the target stars computed from each set of input proper motions were consistent with each other within the uncertainties ($<$ 0.3\,$\sigma$).  In Table~\ref{tab_fgs_results} we present the parallaxes of the target stars derived from using the initial proper motions from UCAC5.

The astrometric fit for the YY Gem field did not include REF05 and REF06 because these stars were fainter than expected and the FGS acquisition failed.  The astrometric fit for the NSVS0103 field did not include REF33; this star was observed in only two out of seven epochs and one of these observations had a very large residual in the fit position.  It was removed from the model so that the solution would converge for the remaining stars.

The FGS parallaxes have errors from the fitting routine in the range of $0.4-0.6$ mas.
This level of performance is slightly worse than that of \citet{Benedict2007AJ....133.1810B} and \citet{vandenBerg2014ApJ...792..110V}, who achieved parallaxes with errors on order $\pm 0.25$ mas.  This is likely due to the limitations on the available {\it HST} roll angles and the selection of background reference stars as described in \S \ref{sec_HSTobservations}.

In addition to the five M-dwarf eclipsing binaries, we found that two of the reference stars (NSVS0103-REF68 and CM-DRA-REF47) are at similar distances and form a common proper motion pair with the affiliated target star.  We solved for the parallax of both stars from the FGS data.  For NSVS0103-REF68 we also kept the proper motion as a free parameter, but could not do the same for CM-DRA-REF47.  With only two observation dates that included CM-DRA-REF47, it was necessary to keep the input proper motion as a constraint in the astrometric grid. These stars are discussed in more detail in Section~\ref{sec_serendipity}.

Based on these trigonometric parallaxes, the derived parameters for our target systems (mass, luminosity, and linear radius for each component) are presented in Table \ref{tab_binaryMLR}, incorporating the $F_{\rm BOL}$ and $\theta$ {\tt binarySED} fit parameters noted in Table \ref{tab_binarySEDdata}.

\begin{deluxetable*}{lrrr}
\tablecaption{FGS Parallaxes \label{tab_fgs_results}}
\tablewidth{0pt}
\tablehead{
\colhead{Star} & \colhead{FGS - UCAC5} & \colhead{FGS -  {\it Gaia}} & \colhead{ {\it Gaia} DR2} \\
    \colhead{} & \colhead{$\pi$ (mas)} & \colhead{$\pi$ (mas)} & \colhead{$\pi$ (mas)} }                
\startdata
        CM-DRA & 68.62 $\pm$ 0.44 & 68.23 $\pm$ 0.38 & 67.34 $\pm$ 0.05\\
  CM-DRA-REF47 & 66.15 $\pm$ 2.57 & 65.10 $\pm$ 1.40 & 67.32 $\pm$ 0.02\\
        GU-BOO &  3.18 $\pm$ 0.59 &  3.15 $\pm$ 0.56 &  6.15 $\pm$ 0.02\\
      NSVS0103 & 15.13 $\pm$ 0.57 & 14.92 $\pm$ 0.53 & 16.48 $\pm$ 0.03\\
NSVS0103-REF68 & 14.98 $\pm$ 0.70 & 14.84 $\pm$ 0.66 & 16.57 $\pm$ 0.05\\
     TRES-HER0 &  5.89 $\pm$ 0.57 &  5.58 $\pm$ 0.53 &  7.11 $\pm$ 0.04\\
        YY-GEM & 67.34 $\pm$ 0.43 & 67.22 $\pm$ 0.40 & 66.23 $\pm$ 0.05\\
\enddata 
\tablecomments{The FGS astrometric fits in the column labeled ``FGS - UCAC5" use the UCAC5 proper motions and the spectrophotometric distance estimates as inputs for the reference stars (Section~\ref{sec_trig_parallax}).  The FGS astrometric fits in the column labeled ``FGS -  {\it Gaia}'' use the  {\it Gaia} DR2 proper motions and parallaxes as inputs for the reference stars (Section~\ref{sec_compare_gaia}).  The last column lists the parallaxes directly from the  {\it Gaia}  DR2 catalog for comparison.}
\end{deluxetable*}




\begin{deluxetable*}{l|ccc|ccc|}




\tablecaption{Primary and secondary derived parameters.\label{tab_binaryMLR}}


\tablehead{\colhead{}  & \multicolumn{3}{|c|}{Primary}& \multicolumn{3}{|c|}{Secondary}\\
\colhead{Target} & \colhead{M} & \colhead{L} & \colhead{R}& \colhead{M} & \colhead{L} & \colhead{R} }  
\startdata
TrES-Her0 & $0.493 \pm 0.003$ & $0.037 \pm 0.006$ & $0.552 \pm 0.058$ & $0.489 \pm 0.003$ & $0.039 \pm 0.006$ & $0.600 \pm 0.063$ \\
CM Dra & $0.231 \pm 0.001$ & $0.0057 \pm 0.0001$ & $0.244 \pm 0.017$ & $0.214 \pm 0.001$ & $0.0050 \pm 0.0001$ & $0.228 \pm 0.016$ \\
GU Boo & $0.610 \pm 0.007$ & $0.345 \pm 0.130$ & $1.289 \pm 0.248$ & $0.599 \pm 0.006$ & $0.334 \pm 0.130$ & $1.333 \pm 0.256$ \\
NSVS0103 & $0.543 \pm 0.003$ & $0.048 \pm 0.005$ & $0.638 \pm 0.053$ & $0.498 \pm 0.003$ & $0.037 \pm 0.004$ & $0.601 \pm 0.054$ \\
YY Gem & $0.599 \pm 0.005$ & $0.073 \pm 0.001$ & $0.658 \pm 0.047$ & $0.599 \pm 0.005$ & $0.066 \pm 0.001$ & $0.624 \pm 0.043$ \\
\enddata
\tablecomments{Mass, luminosity, and linear radius for our target systems, as discussed in \S\ref{sec_trig_parallax}.\\}




\end{deluxetable*}

\subsection{Comparison with {\it Gaia} DR2 Parallaxes} \label{sec_compare_gaia}

Comparisons of our results with the {\it Gaia} DR2 catalog \citep{Gaia2018A&A...616A...1G} can be performed in two ways: (1) using the {\it Gaia} data as input to our astrometric fit, or (2) directly comparing {\it Gaia} parallax values with our results for the target stars.

\begin{figure}
\plotone{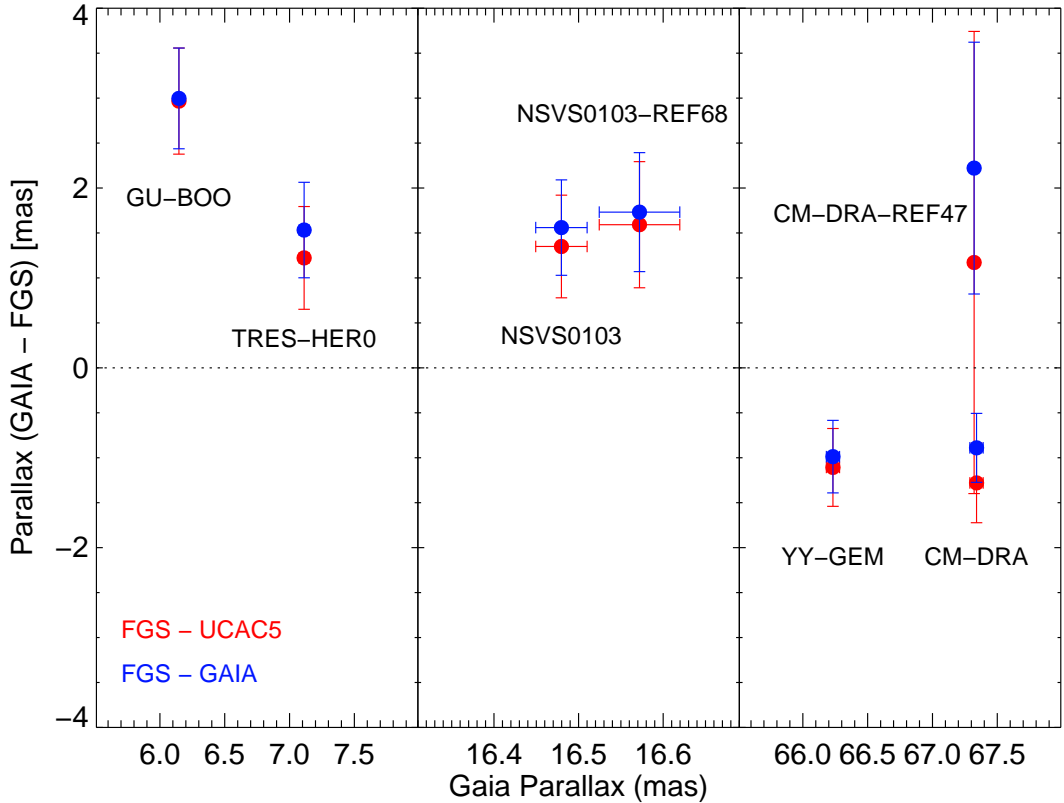}
\caption{\label{fig_par_fgs_gaia} Comparison of FGS parallaxes with {\it Gaia} DR2 parallaxes. The red symbols represent residuals from the FGS parallaxes determined from the input spectrophotometric distances and the UCAC5 proper motions while the blue sumbols represent residuals from the FGS parallaxes determined from the input reference star parallaxes and proper motions from {\it Gaia} DR2. For more detail, see Section \ref{sec_compare_gaia}.}
\end{figure}

As a check on how the reference star distance estimates impacted the final parallaxes of the target stars, we performed an astrometric fit using the parallaxes and proper motions from the {\it Gaia} DR2 catalog as the initial parameters for the reference stars.  The results are presented in the middle column in Table~\ref{tab_fgs_results}.  The last column of Table~\ref{tab_fgs_results} lists the parallaxes of the target stars directly from the {\it Gaia} DR2 catalog.  A comparison of the two FGS astrometric fits (one using UCAC5 proper motions and the spectrophotometric distance estimates as inputs; the other using the {\it Gaia} DR2 proper motions and parallaxes as inputs for the reference stars) with the {\it Gaia} DR2 parallaxes is shown in Figure~\ref{fig_par_fgs_gaia}.  The parallaxes derived from the FGS data consistently produce similar parallaxes, whether using the input values from {\it Gaia}, UCAC5, or one of the other input catalogs.  However, the actual {\it Gaia} parallaxes for the target stars are 2\,$\sigma$ to 5\,$\sigma$ discrepant from the FGS fits.  We are not able to give a definite reason for this discrepancy, but under the assumption that {\it Gaia} parallax values present a ``ground truth" of sorts, we suspect that there may be systematic errors associated with the FGS results, possibly caused by the limited availability of {\it HST} roll angles in two gyro mode. This may imply that the exact same set of astrometric reference stars could not be observed during every epoch (illustrated in Figure \ref{fig_pickle_plots}) which might have impacted the astrometric model used to translate each field of view into the global reference frame.


\section{Common Proper Motion Pairs} \label{sec_serendipity}

The unanticipated presence of two common proper motion (CPM) companions amongst our reference stars -- REF68 for NSVS103, and REF47 for CM Dra -- prompted us to seek additional evidence of CPM companions.  For the purposes of this paper, our criteria for qualifying as a possible CPM companion was merely agreement at the 3-$\sigma$ level between parallax, and proper motion in both right ascension and declination; these possible CPM companions as indicated by {\it Gaia} DR2 data are listed in Table \ref{tab_CPM}.  ({\it HST}/FGS data were not used for CPM identification since it did not conduct a complete survey of all stars in the vicinity of our target stars.)  For YY~Gem, its membership in the Castor sextuple system has already been well-established \citep[e.g.,][and references therein]{Torres2002ApJ...567.1140T}.  For TrES-HER0 and GU~Boo, we examined the now-available {\it Gaia} DR2 dataset for CPM companions; for the former we found one likely companion, and for the latter, some possibilities.




\begin{deluxetable*}{llrrrrrll}




\tablecaption{Common proper motion (CPM) companions and possible CPM companions for the eclipsing binaries in our study, as indicated by {\it Gaia} DR2 data.\label{tab_CPM}}


\tablehead{\colhead{Target} & \colhead{Gaia ID} & \colhead{Sep} & \colhead{Sep} & \colhead{Plx} & \colhead{pmRA} & \colhead{pmDE} & \colhead{typing} & \colhead{Notes} \\
\colhead{} & \colhead{} & \colhead{(arcsec)} & \colhead{(AU)} & \colhead{(mas)} & \colhead{(mas)} & \colhead{(mas)} & \colhead{} & \colhead{} \\
}  
\startdata
TrES-Her0 & 1407718450873494784 &  &  & $7.113 \pm 0.038$ & $-12.075 \pm 0.070$ & $29.904 \pm 0.088$ &  &  \\
 & 1407718450873494528  & 8.2 & 1148 & $7.196 \pm 0.069$ & $-12.959 \pm 0.126$ & $30.299 \pm 0.158$ & $\sim$M5.5V &  \\
NSVS0103 & 1715299716278321408 &  &  & $16.480 \pm 0.030$ & $-106.773 \pm 0.063$ & $58.861 \pm 0.051$ &  &  \\
 & 1715287999607537408 & 65.3 & 3963 & $16.572 \pm 0.048$ & $-107.709 \pm 0.099$ & $59.721 \pm 0.080$ & M4.5V & REF68 \\
CM Dra & 1431176943768690816 &  &  & $67.340 \pm 0.051$ & $-1113.612 \pm 0.114$ & $1181.211 \pm 0.100$ &  &  \\
 & 1431176943768691328 & 26.5 & 394 & $67.322 \pm 0.023$ & $-1110.868 \pm 0.048$ & $1199.405 \pm 0.043$ & WD & REF47 \\
GU Boo & 1278589709364139520 &  &  & $6.147 \pm 0.016$ & $22.728 \pm 0.022$ & $-30.809 \pm 0.031$ &  &  \\
 & 1278514938276783360 & 2106.8 & 342563 & $6.871 \pm 0.755$ & $22.061 \pm 1.345$ & $-30.104 \pm 1.574$ & unknown & Uncertain CPM \\
 & 1278689215165895552 & 1047.1 & 170256 & $6.111 \pm 1.294$ & $24.238 \pm 2.288$ & $-38.721 \pm 2.327$ & unknown & Uncertain CPM \\
 & 1278753639675232128 & 769.2 & 125080 & $5.956 \pm 0.058$ & $28.329 \pm 0.073$ & $-38.046 \pm 0.104$ & unknown & Uncertain CPM \\
YY Gem & 892348454394856064 &  &  & $66.232 \pm 0.051$ & $-201.490 \pm 0.087$ & $-97.104 \pm 0.074$ &  &  \\
 & No data on Castor  & 70.4 & 1063 &  &  &  & A1V/? +  &  \\
 & (Gaia bright limit) &   &   &  &  &  & A2V/? & 
\enddata

\tablecomments{For more details, see Section \ref{sec_serendipity}.}




\end{deluxetable*}

\clearpage


\subsection{NSVS0103: M Dwarf} \label{sec_mdwarf}

NSVS0103-REF68 forms a common proper motion pair with NSVS0103, at an angular distance of 65$\farcs$3.  Based on HYDRA spectra for this object, REF68 is a M4.5V star (Fig.~\ref{fig_serendipity}).  The FGS results place the two stars at a distance of around 67 pc, while {\it Gaia} DR2 places them at a slightly closer distance of 60.7 pc. From the {\it HST}/FGS distance, the linear separation between NSVS0103 and REF68 is 4,400 AU.




\subsection{CM Dra: White Dwarf} \label{sec_whitedwarf}

\begin{figure}
\epsscale{1.00}
\plotone{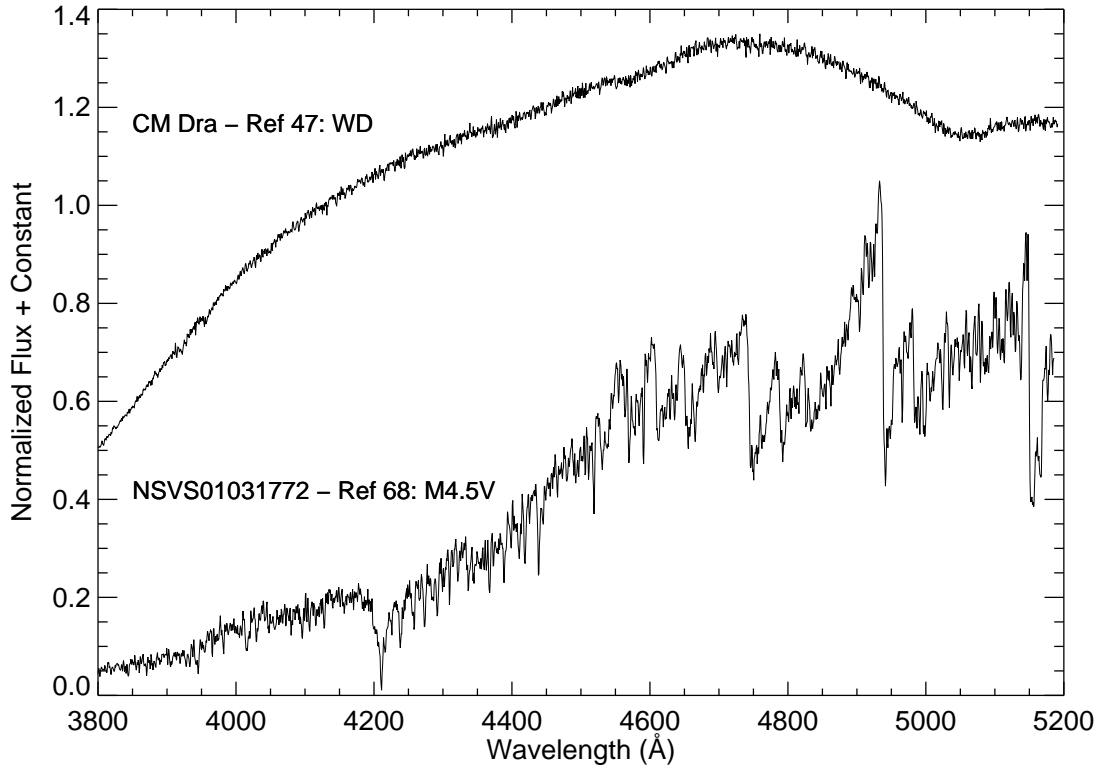}
\caption{\label{fig_serendipity} WIYN Hydra spectra showing that REF47 for CM Dra is a white dwarf (Section \ref{sec_whitedwarf}), and REF 68 for NSVS0103 has a spectral type of M4.5V (Section \ref{sec_mdwarf}). The spectra have not been flattened to remove the spectral continuum nor the instrument response.}
\end{figure}

REF47 associated with CM Dra was previously identified as a white dwarf with a common proper motion \citep{Eggen1967ApJ...148..911E, Greenstein1969ApJ...158..281G}, however, we did not realize this when selecting reference stars in the field.  The HYDRA data show a nearly featureless spectrum (Fig.~\ref{fig_serendipity}), consistent with a white dwarf.  At the {\it HST}/FGS distance of 14.6 pc to CM Dra, its separation of 26$\farcs$5 corresponds to a linear distance of 390 AU.  An increasing number of compact companions are being discovered near main sequence stars \citep[e.g., ][]{Kane2019ApJ...875...74K}, a population to which this object appears to belong.




\subsection{Other CPM data} \label{sec_other_CPMs}

The {\it Gaia} data relating to CPM data for the previous two stars and our other 3 eclipsing binaries can be found in Table \ref{tab_CPM} \citep{Gaia2016A&A...595A...1G,Gaia2018A&A...616A...1G}.

YY Gem's membership as part of Castor sextuple system is as the lowest mass binary on the outskirts of a much higher mass pair of A-type main sequence stars, each with their own lower mass secondaries.  At an angular separation of 70$\farcs$4, at a system distance of 15.1pc this corresponds to a linear distance of 1,060 AU from the close quadruple Castor Aab+Bab.

TrES-Her0 has a CPM companion at 8$\farcs$2 corresponding to a linear separation 1,150 AU.  The nature of that companion is unknown; however, with $\Delta$G = 1.8mag, its spectral type will be roughly M5.5V.

The {\it Gaia} DR2 for GU Boo does not definitively indicate one or more CPM companions, given the error bars parallax and proper motion for 3 possible CPM candidates.  However, these candidates are possibly nearby in $\{ \pi, {\rm PM}_{\rm RA}, {\rm PM}_{\rm DE} \}$ space, based upon the large errors in their astrometric solutions; the currently reported DR2 values for these objects indicate linear distances from 125,000 to 350,000 AU.

\subsection{Discussion}

A review of the literature indicates that the presence of CPM pairs to these close binaries is not unexpected.  \citet{Tokovinin2006A&A...450..681T} found a higher order multiplicity fraction of close binaries (e.g. Figure 14 from that study) that indicated most ($\sim$ 63\%) had a tertiary component at larger separations.  There are several possible formation scenarios that have been proposed for how wide multiple systems such as the ones examined above may have formed.  These include the unfolding of triple systems from a protostellar core \citep{Reipurth2012Natur.492..221R}, the gravitational binding of stars as a cluster dissolves \citep{Kouwenhoven2010MNRAS.404.1835K} and the gravitational binding of stars that form in adjacent protostellar cores \citep{Tokovinin2017MNRAS.468.3461T}.  While more information is needed to definitively determine the formation mechanism of these pairs, early indications, based on their separations and multiplicity, seem to suggest that systems may have been the result of the unfolding of a triple system in a protostellar core, in particular CM Dra, NSVS0103 and TrES-Her0.  This scenario would also allow for the range in outer orbit separations that we see in our systems, which range from $\approx 400 - 4,000$ AU and are still within the average protostellar core size of 0.1 pc.  We caution that this is an extremely small subset of wide pairs which does not explain how the vast majority of wide systems are formed.  It is much more likely that wide pairs form through a variety of different mechanisms and it remains to be seen if one dominates over the others.

%
%
%

%
%
%
%
%
%
%


\section{Distance-Enabled Astrophysics} \label{sec_discussion}




The key results from this study are the mass, luminosity, and radius values for these low-mass stars.  We compared our values to similar data products found in \citet{Torres2010A&ARv..18...67T}; this is of particular interest given their approach was separate but complementary (eclipsing binary orbit fitting) to ours (trignometric parallax combined with SED fitting).
Data from the dynamical parallax study of \citet[][B16]{Benedict2016AJ....152..141B} are not used here, since absolute luminosities are not determined in B16, just absolute magnitudes $M_{\rm V}$ and $M_{\rm K}$.  The techniques presented in this study, particularly the use of {\tt binarySED} for determination of component true luminosities, would be readily applicable to the objects studied in B16, but this is beyond the scope of the current study\footnote{This is true in many cases: published `mass-luminosity' relationships \citep[e.g.,][]{Delfosse2000A&A...364..217D}  are actually `mass-absolute magnitude' relationships, making direct comparison to stellar luminosity problematic.  The departures from a true luminosity function for the  mass-absolute magnitude relationships is readily apparent in the scatter of the $V$-band relationship seen in B16's figure 24.}.

\begin{figure*}
\epsscale{1.00}
\plotone{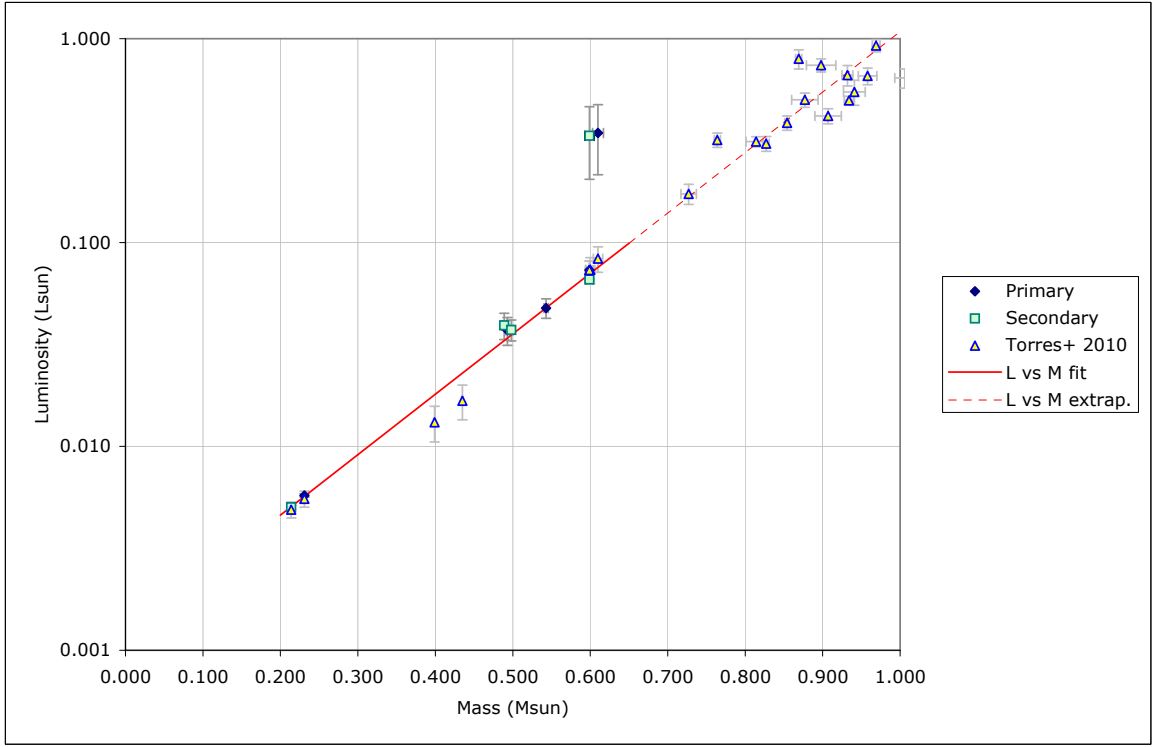}
\caption{\label{fig_LvsM} Luminosity versus mass for our target stars, as well as the corresponding points from \citet{Torres2010A&ARv..18...67T}.  The line fit (Eqn. \ref{eqn-LvsM-FGS}) is restricted to the range 0.20 -- 0.65 $M_\odot$, with an extrapolation of that fit shown to 1.0 $M_\odot$. The outlier pair are the GU Boo components -- they are not used in calculating the fit shown. For more details, see \S \ref{sec_discussion}.}
\end{figure*}

\begin{figure*}
\epsscale{1.00}
\plotone{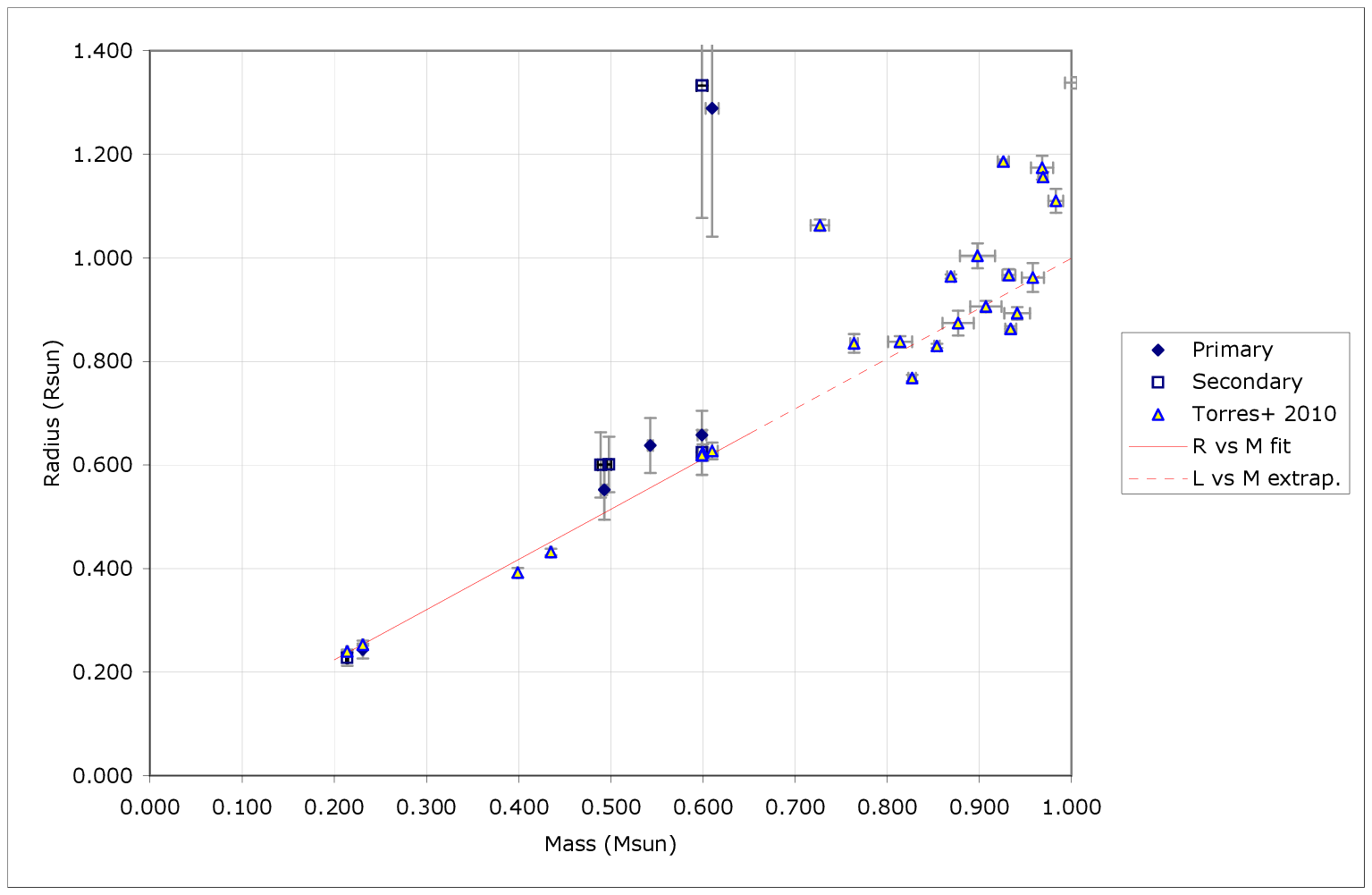}
\caption{\label{fig_RvsM} Radius versus mass for our target stars, as well as the corresponding points from \citet{Torres2010A&ARv..18...67T}.  The line fit (Eqn. \ref{eqn-RvsM-FGS}) is restricted to the range 0.20 -- 0.65 $M_\odot$, with an extrapolation of that fit shown to 1.0 $M_\odot$ -- as in Fig. \ref{fig_LvsM}, the line fit shown does not include the outlier GU Boo. For more details, see \S \ref{sec_discussion}.}
\end{figure*}

For the luminosity versus mass and radius versus mass plots (Figures \ref{fig_LvsM} and \ref{fig_RvsM}, respectively), it is readily apparent that our distance for GU Boo appears to be incorrect: in both cases, our results and the \citet{Torres2010A&ARv..18...67T}, aside from GU Boo, are clearly consistent with the relationship fit lines discussed below.  Comparing our FGS parallax-derived distance to the distance estimate for GU Boo found in \citet{Lopez-Morales2005ApJ...631.1120L}, there is considerable discrepancy -- our FGS value of $313.5 \pm 59.0$ pc versus their estimate of $\sim 140$ pc.  {\it Gaia} DR2 places GU Boo at a distance of 162.6 $\pm$ 0.5 pc.  Testing the \citet{Lopez-Morales2005ApJ...631.1120L} distance in our evaluation of the associated GU Boo parameters of radius and luminosity indicates they converge to values consistent with the fits seen in Figures \ref{fig_LvsM} and \ref{fig_RvsM}.

Examining our FGS data, there is no clear cause for this discrepant value:  There is no particularly problematic astrometric reference star which, when suppressed, causes the astrometric solution to change substantially.  Nor is the target star particularly faint.  However, among all of all our targets, GU Boo did have the faintest reference stars, at an average of 0.25-1.0 magnitudes fainter than the other target reference ensembles.  It is also one of the most distant astrometric targets, though TrES-Her0 is also at 100+ pc in distance.  GU Boo does have the greatest brightness ratio between its two components (Table \ref{tab_ratio_data}), but this should not affect measurements any more than our other targets:  first, the projected angular size of the orbit is small, so astrometric wander due to the bright/dim components is undetectable by Gaia or {\it HST}/FGS.  Second, the photometric variation of being in/out of eclipse is large for all our science targets (roughly a factor of 50\%) and only slightly larger for GU Boo (up to roughly 45\%, depending on bandpass).

For completeness, we have kept this object on these plots; however, for the fits described below, we have excluded it from the fitting process.

{\it Mass versus Luminosity.}  For our target stars, with determined distances $d$ and bolometric fluxes $F_{\rm BOL}$, luminosity is readily established from the familiar relationship $L = 4 \pi d^2 F_{\rm BOL}$,
with $d$ in cm and $F_{\rm BOL}$ in erg cm$^{-2}$ s$^{-1}$.
We can directly relate these numbers to fractional solar luminosity values using a reference value of $3.83 \times 10^{33}$ erg cm$^{-2}$ s$^{-1}$ for the Sun \citep{Cox2000asqu.book.....C}.  These results are presented in Figure \ref{fig_LvsM}.
The resultant luminosity versus mass fit is:
\begin{equation}\label{eqn-LvsM-FGS}
\log L = M \times (2.960 \pm 0.033) -(2.928 \pm 0.012)
\end{equation}
We chose the upper bound of 0.65 $M_\odot$ for our data which corresponds roughly to the expected `turn-on' point, below which stars become fully convective \citep{Chabrier2000ARA&A..38..337C}.
Fitting for the luminosities indicated by the {\it Gaia}
parallaxes (including GU Boo), the fit in Equation \ref{eqn-LvsM-FGS} becomes:
\begin{equation}\label{eqn-LvsM-Gaia}
\log L = M \times (2.920 \pm 0.023) -(2.915 \pm 0.009)
\end{equation}


This is within 1-$\sigma$ of the idealized $L \propto M^3$ relationship for this mass range \citep{Eddington1924MNRAS..84..308E,Eddington1925MNRAS..85..403E}.

{\it Radius versus Luminosity.}  Combining distance with estimates of angular size from the SED fitting, linear size is readily inferred.  For the sake of clarity, it is worth emphasizing these angular size estimates are simply estimates: the template that the SED fitting process selects as the best-fit template for the input photometry will have an {\it a priori}, fixed effective temperature ($T_{\rm EFF}$) value associated with it.  From the measured $F_{\rm BOL}$ and that $T_{\rm EFF}$, an angular size can be estimated.  This is less direct of an approach to obtaining linear radius than an eclipsing binary orbit solution technique, but it is complementary.  These data are presented in Figure \ref{fig_RvsM}; as with the previous figure, included in this figure is a fit for stars from our study, excluding GU Boo:
\begin{equation}\label{eqn-RvsM-FGS}
R = M \times (0.970 \pm 0.012) - (0.030 \pm 0.004)
\end{equation}
As seen in Figure \ref{fig_RvsM}, there are some $\sim 1\sigma$ outliers in radius, in the $M=0.5-0.6 M_\odot$ range; this mass range has been problematic for angular size estimates and is currently an area of active research \citep[e.g.][]{Mann2015ApJ...804...64M}.  As above in Equation \ref{eqn-LvsM-Gaia}, a similar fit was established for our program stars, but using radii derived using {\it Gaia} parallaxes, inclusive of GU Boo:
\begin{equation}\label{eqn-RvsM-Gaia}
R = M \times (1.103 \pm 0.050) - (0.008 \pm 0.021)
\end{equation}
This is within 2.5-$\sigma$ of the idealized $R \propto M$ relationship for this mass range \citep{Eddington1924MNRAS..84..308E,Eddington1925MNRAS..85..403E}

The eclipse binary timing of \citet{Torres2010A&ARv..18...67T} provides better precision on radius than our approach -- and importantly, is a direct rather than indirect measure of radius.  However, for those stars (especially single stars) where the system is not eclipsing, this study demonstrates a path forward in at least indirectly assessing the stellar sizes.  This is especially appealing in the {\it Gaia} DR2 era, where high-precision distances will be available for these objects.

Examining the $R$ versus $M$ fit line extrapolation out to 1 $M_\odot$, we see that in relationship to the higher mass points from \citet{Torres2010A&ARv..18...67T}, this extrapolation appears to agree with the lower bound of the stars in this radius-mass space.  It is possible that the objects lurking above the line are beginning to evolve off the main sequence, inflating in size.

Finally, it is worth noting that the SED fitting approach of \S \ref{sec_binarysed} will be of general utility for upcoming {\it Gaia} DR2 results.  Binaries observed by {\it Gaia} will probably be close enough for crude orbit fitting and determination of dynamical parallaxes  -- e.g. see the reflex motion plots of \citet{Benedict2016AJ....152..141B} -- though with significant complicating factors to be considered. First, the {\it Gaia} temporal sampling function convolved with the binary orbital periods will possibly result in serious aliasing of the observed astrometric reflex motion.  Second, in some of the unequal brightness cases, while a reflex motion will be detected, the spatial resolution of the instrument is insufficient to spatially separate the two components of a binary, so a robust estimate of that brightness ratio in the {\it Gaia} observation bandpass will be necessary.  A tool like {\tt binarySED} will be essential to addressing at least this second concern, and will also be essential for equal brightness ratio binaries.



%
%
%
%
%
%
%
%
%
%

\section{Conclusion}\label{sec_conclusion}


For this work, we measured trigonometric parallaxes to five eclipsing binary M-dwarf / M-dwarf systems. By directly determining their distances and using SED fitting to measure their bolometric fluxes, we are able to provide semi-empirical constraints on their physical properties, which are in good agreement with the complementary eclipsing binary values from \citet{Torres2010A&ARv..18...67T}.  Our approach of distance-leveraged SED fitting will be of more general utility for measuring luminosities and inferring linear radii from high-precision distances coming from {\it Gaia} since the vast majority of its observed stars will be non-eclipsing binaries or single stars.
Finally, provided that robust determination of angular sizes from SED fitting can be made at the 1\% level with matching distance precision from {\it Gaia}, linear radius discrepancies reported for M-dwarf stars \citep[e.g.,][]{Berger2006ApJ...644..475B,Lopez-Morales2007ApJ...660..732L,Lubin2017ApJ...844..134L} can be straightforwardly investigated with statistically significant samples.  These large samples could be correlated with indicators of magnetic activity, rotation rates, multiplicity / proximity of companions, and other phenomena proposed to account for discrepancies in linear radii.




\acknowledgements


We would like to thank Andy Boden of Caltech for his development of the {\tt sedFit} software suite, which was instrumental in deriving these results.  We would also like to acknowledge our anonymous referee, who made many positive suggestions for improving this manuscript.
Support for this work was provided by NASA through grant GO-11213 from the Space Telescope Science Institute, which is operated by the Association of Universities for Research in Astronomy, Inc., under NASA contract NAS5-26555.
This publication makes use of data products from the Two-Micron All Sky Survey, which is a joint project of the University of Massachusetts and the Infrared Processing and Analysis Center/Caltech, funded by NASA and the NSF. This research has utilized the SIMBAD database and VizieR catalogue access tool,
both operated at CDS, Strasbourg, France \citep{Wenger2000A&AS..143....9W,Ochsenbein2000A&AS..143...23O}, and the NASA's Astrophysics Data System Bibliographic Services.

{\it Facilities}: {\it HST}-FGS, {\it Gaia}, WIYN-HYDRA, Lowell: 31-inch USGS telescope, 42-inch Hall telescope

{\it Software}: {\tt sedFit}, {\tt GaussFit}, {\tt IRAF DOHYDRA}

{\it ORCID iDs}:\\
GTvB \url{https://orcid.org/0000-0002-8552-158X}; \\
GHS \url{https://orcid.org/0000-0001-5415-9189}; \\
KvB \url{https://orcid.org/0000-0002-5823-4630}; \\
ZH \url{https://orcid.org/0000-0003-4236-6927}; \\
TSB \url{https://orcid.org/0000-0001-9879-9313}


\bibliographystyle{apj}
\bibliography{journal-references}


\clearpage

\end{document}